\documentclass[manuscript]{aastex}

\newcommand{\jb}{\hbox{\sl J}}
\newcommand{\hb}{\hbox{\sl H}}
\newcommand{\kb}{\hbox{\sl K{$_S$}}}

\slugcomment{To appear in Astronomical Journal in 2009}

\shorttitle{Infrared Photometry of the ONC}
\shortauthors{Robberto et al.}

\usepackage{amssymb}

\begin{document}


\title{A\ Wide-Field Survey of the Orion Nebula Cluster in the Near-Infrared}


\author{M. Robberto, D. R. Soderblom}
\affil{Space Telescope Science Institute, Baltimore, MD 21218}
\email{robberto@stsci.edu,drs@stsci.edu}

\author{G. Scandariato}
\affil{\mbox{Dipartimento di Fisica e Astronomia, Universit\`a di Catania, Italy}} 
\email{gas@oact.inaf.it}

\author{K. Smith, N. Da Rio} 
\affil{Max-Planck-Institut f\"ur Astronomie, Heidelberg, Germany}
\email{smith@mpia-hd.mpg.de,dario@mpia-hd.mpg.de}

\and

\author{I.\ Pagano, L.\ Spezzi\altaffilmark{1}}
\affil{INAF -- Osservatorio Astrofisico di Catania, Italy.}
\email{ipa@oact.inaf.it, lspezzi@oact.inaf.it}

%\email{aastex-help@aas.org}

%\author{various others\altaffilmark{5}}


\altaffiltext{1}{current address:\ European Space Agency (ESTEC), PO\ Box 299, 2200 AG\ Noordwijk, The Netherlands}
% \altaffiltext{2}{Society of Fellows, Harvard University.}
% \altaffiltext{3}{present address: Center for Astrophysics,
%     60 Garden Street, Cambridge, MA 02138}
% \altaffiltext{4}{Visiting Programmer, Space Telescope Science Institute}
% \altaffiltext{5}{Patron, Alonso's Bar and Grill}


\begin{abstract} 

We present \jb, \hb\ and \kb\ photometry of the Orion Nebula Cluster obtained at the CTIO/Blanco 4~m telescope in Cerro Tololo with the ISPI imager. From the observations we have  assembled a catalog of  about $\sim7800$ sources distributed over an area of approximately $30\arcmin\times40\arcmin$,   the largest of any survey deeper than 2MASS in this region. The catalog provides absolute coordinates accurate to about 0.15 arcseconds and $3\sigma$ photometry in the 2MASS  system down to $\jb\simeq 19.5$~mag,  $\hb\simeq18.0$~mag, $\kb \simeq18.5$~mag, enough to detect planetary size objects 1~Myr old under $A_V\simeq10$~mag of extinction at the distance of the Orion Nebula. We present a preliminary analysis of the catalog, done comparing the (\jb-\hb,\hb-\kb) color-color diagram,  the (\hb,\jb-\hb) and (\kb,\hb-\kb) color-magnitude diagrams and the \jb\hb\kb\ luminosity functions of three  regions at increasing projected distance from the Trapezium. Sources in the inner region typically show IR\ colors compatible with reddened T\ Tauri stars, whereas the outer fields are dominated by field stars seen through an amount of extinction  which decreases with the  distance from the  center.
The color-magnitude diagrams make it possible to clearly distinguish between the main ONC population, spread across the full field,  and background sources. %The position of the ONC\ population peak in the color-magnitude diagrams drifts to lower mass and bluer colors  moving from the inner to the outer region, suggesting that the inner cluster contains a higher fraction of massive and young stars. This may point to a scenario in which star formation in the ONC has proceeded from the outside to the inside while increasing the fficiency of massive star  formation. 
  The luminosity functions of the inner region, corrected for completeness, remain relatively flat  in the sub-stellar regime regardless of the strategy adopted to remove background contamination.  
\end{abstract}



\section{Introduction}\label{sec:intro}
The Orion Nebula hosts the richest cluster of\  young ($\tau\simeq 1$~Myr) Pre-Main-Sequence (PMS) stars within 1~kpc of the Sun and therefore represents an ideal laboratory to understand the process of star formation \citep[see][for recent reviews]{Muench+08, ODell+08}. The cluster, both rich ($n\approx 2000$ members) and dense (about  $2\times10^4$\ sources per cubic parsec at its center), is dominated by a small number of massive OB\ stars mostly clustered in the Trapezium  multiplet ($\theta^1$~Ori). Their UV\ emission has created a blister HII\  \ region ({\sl Orion Nebula, M~42, NGC~1976}) whose ionization front is still carving the underlying Orion Molecular Cloud \cite[OMC-1,][] {ODell+08,ODell+09}. About half of the young cluster members, surrounded by their original circumstellar disks, have been already exposed to the  hard-UV\ radiation of the Trapezium stars, whereas the other half remain enshrouded within the OMC-1, together with newer active sites of  star formation.

Visible and near-IR\ data, both in imaging and spectroscopy, are needed to characterize the main physical parameters of the cluster population \citep{LAH97}. These observations, however, are hampered by the brightness and non-uniformity of the nebular background. To overcome these difficulties, we have exploited the unique combination of sensitivity and spatial resolution offered by  the Hubble Space Telescope (HST)\ to obtain accurate photometry of the cluster, especially at substellar masses (HST\ Treasury program GO-10246). Our extensive HST\ survey has been complemented by ground-based observations, imaging the Orion Nebula from the {\sl U}-band to the  \kb-band at La Silla and Cerro Tololo (CTIO)\ observatories. The ground-based observations, carried out in parallel on the same nights (but at a different epoch than the HST observations), complement the deep HST data which saturate at relatively low brightness levels. Their simultaneity makes the derived stellar colors largely immune to the uncertainties associated with source variability. 

%Moreover, infrared observations allow not only to probe the stellar population still embedded in the molecular cloud, but are also sensitive to the thermal emission reprocessed by the circumstellar disks.

The ground-based  survey at visible wavelengths, performed with the Wide Field Camera (WFI) at the ESO/MPG 2.2~m telescope at La Silla, has been recently presented in \cite{DaRio+09}. In this paper we present the ground-based IR\ survey, performed at the CTIO/Blanco telescope with the ISPI\ imager.
Its combination of sensitivity and field coverage (about $30\arcmin\times40\arcmin$) ideally complements the previous surveys of this region. In particular, it reaches fainter magnitudes than the wide field survey ($30\arcmin\times45\arcmin$) by Hillenbrand et al. (1998), while it covers a much larger area than the deep surveys done with Keck (\hb\kb-bands, $5\farcm1 \times 5\farcm1$, Hillenbrand \&\ Carpenter, 2000),  HST-NICMOS\ (\jb\hb-bands, $2\farcm3\times2\farcm3$, Luhman et al. 2000),  UKIRT\ ({\sl I}\jb\hb-bands,\ 36 arcmin$^2$, Lucas \&\ Roche 2000),  NTT\ (\jb\hb\kb, $5\arcmin\times5\arcmin$,   Muench et al. 2002) and  Gemini (\jb\hb\kb-bands, 26~arcmin$^2$, Lucas et al. 2005). All these deep surveys concentrate on a field centered around  the Trapezium stars, or in its immediate surroundings in the case of \cite{LR00}. 

 In Section\ \ref{sec:obs} we discuss the  observing strategy, while in Section\ \ref{sec:reduction} we describe the data reduction and photometric calibration. In Section\ 4 we present our completeness analysis. The resulting source catalog is presented in Section 5, whereas in Section 6 we discuss the color-color and color-magnitude diagrams and the luminosity function derived from our \jb, \hb, and \kb-band photometry.

\section{ISPI observations}\label{sec:obs}
The Infrared Side Port Imager (ISPI) is the facility infrared camera at the CTIO Blanco 4~m telescope. ISPI uses a 2k $\times$ 2k HgCdTe HAWAII-2 array, with reimaging optics providing a scale of 0.3 arcsec/pixel corresponding to a 10.25 $\times$ 10.25 arcmin field of view. Our target area, about $30\arcmin \times 40\arcmin$, covers the field imaged with the HST. 

The observations were performed  on the nights of 1 and 2 January 2005 (indicated hereafter as Night A and Night B respectively) in the \jb, \hb\ and \kb\ filters using the Double Correlated Sampling readout mode. 
%i.e. immediately after the reset and at the end of the integration.
The main filter parameters are listed in Table~\ref{Tab_ISPI_Filters}. The seeing was $\simeq0\farcs7$ in most of the \kb-band images and occasionally worse on night B.

\begin{figure}
 \epsscale{1}
 \plotone{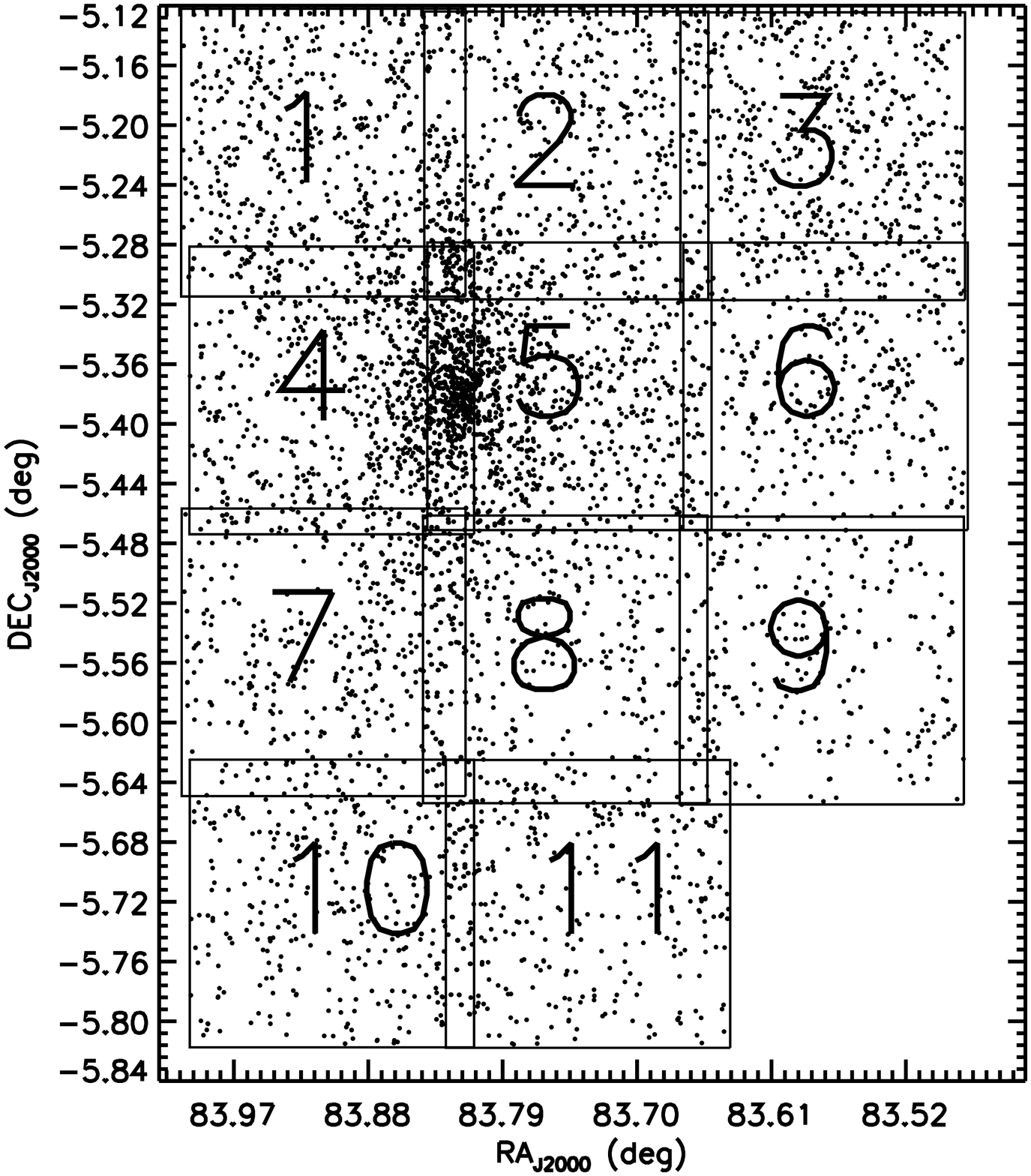}
 \caption{Position of the eleven ISPI fields, superimposed to the positions of the sources detected in our survey.}\label{fig:ISPI_fields}
\end{figure}

The survey area was divided into eleven fields (Figure~\ref{fig:ISPI_fields}). %Field 1 to 6 were observed on night A, whereas field 4 to 11 were observed %on night B. Fields 4, 5, and 6 have therefore been observed on both nights.
The central position of each field, together with the airmass of the observations, is listed in Table~\ref{Tab_ISPI-logbook}. 
Each field was observed with an ABBA\ pattern. After pointing the telescope to the center of a field (A\ position) a   5 point dithering pattern was executed offsetting the telescope approximately by $\pm30\arcsec$ along the field diagonals.
This first group of 5 dithered exposures was followed by a second group of sky frames (B position)\  taken on a nearby field free, or nearly free, of diffuse nebular emission to extend the field coverage and monitor the  flat-field response  close in time to our observations. The sky-source sequence was then repeated back, completing the ABBA cycle.  This provides a total of 10 frames of 30~s exposure for the common part of each field, totalling 300~s integration. Due to the high background, we split the \kb-band 30~s exposures in two consecutive 15~s exposures,  maintaining the background well within the linear regime. 

Each dithering sequence was repeated with a 3~s integration time. These short frames were  coadded into another image of   30~s total integration time, which we used  to extract the photometry of sources saturated in the longer exposures.  Due to an error in our observing procedure, the \jb-band observations of field 3 were obtained only with 3~s exposures.

For absolute calibration we observed at various airmass the faint IR standard stars 9108, 9118 and 9133 of \citet{Persson+1998}. Unfortunately, on both nights the atmospheric transmission at IR\ wavelengths turned out to be unstable due to variations in the water vapor opacity, and we eventually derived the zero point calibration of our images through comparison with the extensive set of 2MASS data across our wide field area, as described in Section\ \ref{sec:calib}.

\section{Analysis}\label{sec:reduction}

The first step in the data reduction was the correction for the intrinsic non-linearity of the detector, which we performed by applying the correction curve kindly provided by N. Van der Bliek\footnote{now also available available on the ISPI web page:\ http://www.ctio.noao.edu/instruments/ir\_instruments/ispi/}. We flat-fielded each image using dome flats and correct for residual color terms at low-spatial frequency using delta-flats derived from  median averaged sky images. The dome flats were also used to flag bad pixels. The  sky frames, properly filtered to remove spurious sources arising from the latent images of bright stars on the infrared detectors, were combined and subtracted from the science images.

Since ISPI exhibits significant field distortion, each science image had to be geometrically corrected before being combined in a dithering group. This required a first pass of aperture photometry with DAOPHOT, in order to extract a source catalog with well measured centroids and photometry (to disentangle the case of multiple candidates in the search radius). These sources were then cross-identified  with the 2MASS catalog for astrometric reference, building an image distortion map relating their position on the original images to the actual position provided by 2MASS. 
%The average shift of each individual images vs.\ 2MASS was then subtracted %and a distortion map, , was created for each image. 
Due to the non-uniform distribution of sources, clustered at the center of the nebula, the accuracy of the distortion correction map varies across each field. Therefore, we combined the distortion maps relative to all images taken on  the same night  to derive, for each filter, a master distortion map. Each individual image was then warped using a 4th-order polynomial fit with coefficients derived from the corresponding distortion map. 
%(with the IDL POLYWARP.pro routine). The warping of the images was then %performed with the IDL POLY\_2D.pro routine. 
Images belonging to the same dithered groups were then combined per integer pixels into single $2500\times2500$  images. The DAOPHOT source extraction was then repeated on these geometrically corrected images to produce a new source catalog from which we derived the final astrometric solution
%information on the fits header was derived 
of each image by minimizing the residual shift and rotation with respect to the 2MASS positions. The average scatter between our coordinates and the corresponding 2MASS\ coordinates turns out to be  about 0.15~arcsec, i.e., half of a ISPI\ pixel. The same procedure has been adopted on the 3~s images. 

In conclusion, for each field the
final
data consist of one combined image in each of the 3 filters (\jb, \hb, \kb) and exposure
times (300~s and 30~s), with the exception of fields 4, 5, and 6, which have double
this number of images, and field 3, which was not observed with deep \jb~
images.

\subsection{Instrumental magnitudes}\label{sec:datasec}
The extraction of source photometry was performed on the final combined images, for both short and long exposure times. The Daophot FIND IDL procedure was used to extract an initial list of candidate sources, with rather loose detection thresholds and criteria to include the largest possible number of candidates. This initial list, counting more than 60,000 candidates, has been visually inspected on the original images to clean up artifacts and to preliminarily classify real sources as either point-like or extended. The list was then reduced to 7563 sources, divided between 6305 point-like and 933 diffuse.

%The final catalog consists of 9268  sources, several detected in more than one band, exposure time, or night (see Section~\ref{sec:res}).
For all sources we first derived an aperture magnitude by integrating the flux over a 5 pixel (1.5\ arcsec)\ radius circular aperture, taking sky annuli typically between 10 and 15 pixels (3.0-4.5~arcsec); for 1416 bright sources, saturated in the 30~s images, we used the photometry extracted from the 3~s images.

%Other 174 sources turned out to be saturated also in the 3s images. They are still present in our catalog with their original 2MASS.

We then performed PSF photometry on all point-like sources. Due to the spatial variability of the PSF across the field of view of ISPI,
present even after geometric correction, we divided each image in 9 parts (3$\times$3 squares) allowing for some overlap between adjacent sub-images. The samples of stars in each sub-image, still relatively rich, had a more homogeneous PSF\  allowing us to reliably derive PSF magnitudes. We subtracted the PSFs\ and looked at the residual images searching for faint companion stars previously undetected. We found in this way 325 ``hidden companions''. Finally, we performed a second pass of PSF\ photometry, first by removing from the  list of stars used to derive the PSF\ those having a faint companion, and then deriving also  the photometry of the newly detected faint companions. 
\subsection{Absolute calibration}\label{sec:calib}
As mentioned in Section\ \ref{sec:obs}, the IR\ photometric quality of our nights turned out to be unsatisfactory due to hygrometric variability. We therefore calibrated our ISPI instrumental magnitudes directly to the 2MASS photometric system. The 2MASS catalog provides absolute photometry in the \jb\ (1.25\micron), \hb\ (1.65\micron), and \kb\ (2.17\micron) bands to a limiting magnitude of 15.8, 15.1, and 14.3, respectively, with signal-to-noise ratio greater than 10. 

%\subsection{Bright source calibration}
To estimate the color terms and zero points of each image, we first scaled the instrumental PSF\ magnitudes of point sources measured in the short 30~s images to the long 300~s images, chosen as the reference time because of their source richness. To minimize systematic effects due do atmospheric variations, we did not simply add the nominal  $\Delta mag=2.5$ factor but compared the instrumental magnitudes of well measured stars to derive for each field and filter a mean magnitude offset $\overline{\Delta mag}$  between the short and long images.   By homogenizing in this way the zero points of the short and long images, we add a small uncertainty to the photometric errors of the bright stars measured in the short exposures. 
%The case of \jb-band field 7 is plotted in . 
%Note that errors on magnitudes less than $\sim$11 increase after normalization, %due to error propagation.}\label{fig:magerrzp}

%\subsection{Calibration to 2MASS system}
For each bandpass $\lambda$\ we then evaluated the relations \begin{displaymath}
 mag_{\rm 2MASS}(\lambda)-mag_{\rm ISPI}(\lambda)=ZP_{\lambda}+\epsilon_{\lambda_1,\lambda_2}\cdot c_{\rm ISPI}(\lambda_1,\lambda_2),
\end{displaymath}
where $mag_{\rm 2MASS}$ is the magnitudes in the 2MASS\ photometric system, $mag_{\rm ISPI}$ is the ISPI\ \ instrumental magnitude, the intercept $ZP_{\lambda}$ is the zero point in a given bandpass, the slope $\epsilon_{\lambda_1,\lambda_2}$ is a color coefficient between the wavelengths $\lambda_1$\ and  $\lambda_2$ and $c_{\rm ISPI}$ is the corresponding observed ISPI\ color.
We found the strongest statistical correlation in the combination:
\begin{displaymath}
 H_{\rm 2MASS}-H_{\rm ISPI}=ZP_{H}+\epsilon_{JK}\cdot (J-K)_{\rm ISPI}.
\end{displaymath}
For each field, we derived the linear fit parameters $ZP_H$\ and $\epsilon_{JK}$ using an iterative procedure to rejects spurious outliers and all sources with errors greater than $0.1^{m}$ in the 2MASS catalog and $0.05^{m}$ in our input catalog. As shown in Table \ref{tab:fitpar}, the color coefficient  $\epsilon_{JK}$ varies slightly from field to field during each night, whereas the zeropoint $ZP_\lambda$ shows an increase towards the middle of the night (observation were scheduled with fields 4 and 10 transiting at the meridian, in the two nights).

\begin{figure*}
   \epsscale{1.}
  \plotone{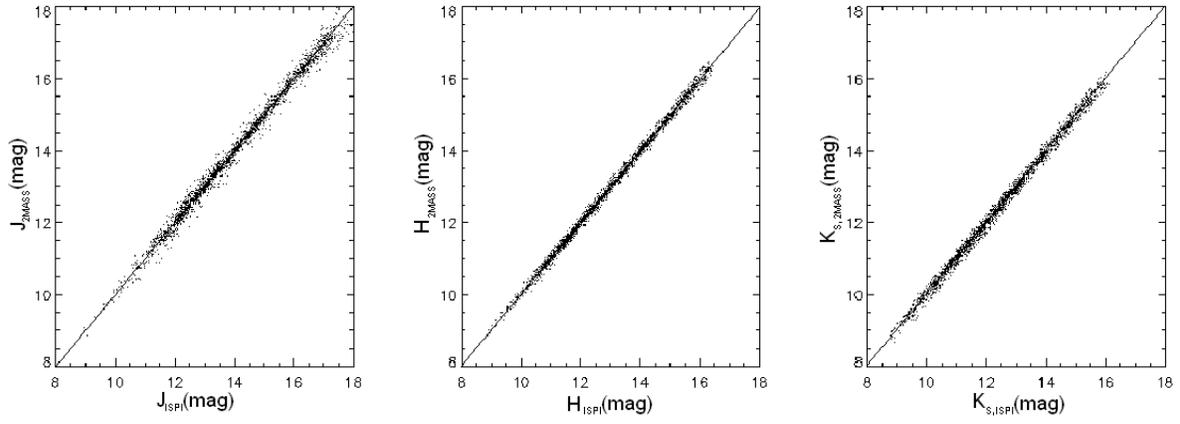}%\\
%   \plotone{lin2lin.eps}
   \caption{Comparison between 2MASS magnitudes and ISPI\ calibrated magnitudes. %(upper panel) and comparison between field 4 and field 5 magnitudes (lower panel). 
%Straight lines in each plot represent $y=x$ function. Sistematic deviation from trends indicated by lines in the lower panel are due to inaccuracies in field distortion correction.
}\label{fig:lin2lin}
 \end{figure*}

For the \jb\ and \kb\ bands, we derived the zero-points and the color terms from a linear fit to the color relations
\begin{eqnarray}
 &(J-H)_{\rm 2MASS}=ZP_{JH}+\epsilon_{JH}\cdot (J-H)_{\rm ISPI},\nonumber\\
 &(H-K_{S})_{\rm 2MASS}=ZP_{HK}+\epsilon_{HK}\cdot (H-K_{S})_{\rm ISPI}.\nonumber
\end{eqnarray}
This allowed us to derive magnitudes calibrated   in the 2MASS\ system for all  sources with ISPI $JHK$ instrumental photometry. In Figure \ref{fig:lin2lin} we show a comparison between the magnitudes reported in the 2MASS catalog and our calibrated magnitudes.
%; we also plot in Figure \ref{fig:lin2lin} a comparison between magnitudes of objects shared by field 4 and field 5. 
The plots show the lack of systematic differences, besides the random errors that we attribute to photometric errors or stellar variability.

For sources lacking one or two magnitudes (typically either \jb\ or both \jb\ and \hb), we assumed a linear relation of this type
\begin{displaymath}
 mag_{\rm 2MASS}(\lambda)=\alpha\times mag_{\rm ISPI}(\lambda)+\beta
 \end{displaymath}
deriving $\alpha$ and $\beta$, for each bandpass and  field, from the sample of calibrated sources.   The parameter $\alpha$   provides a magnitude-dependent correction to the zero point $\beta$, which is appropriate since fainter sources have, on average,  redder colors. Using these coefficients we calibrated the magnitudes, with the relative uncertainties, of all remaining sources.

\begin{figure*}
 \epsscale{1}
 \plotone{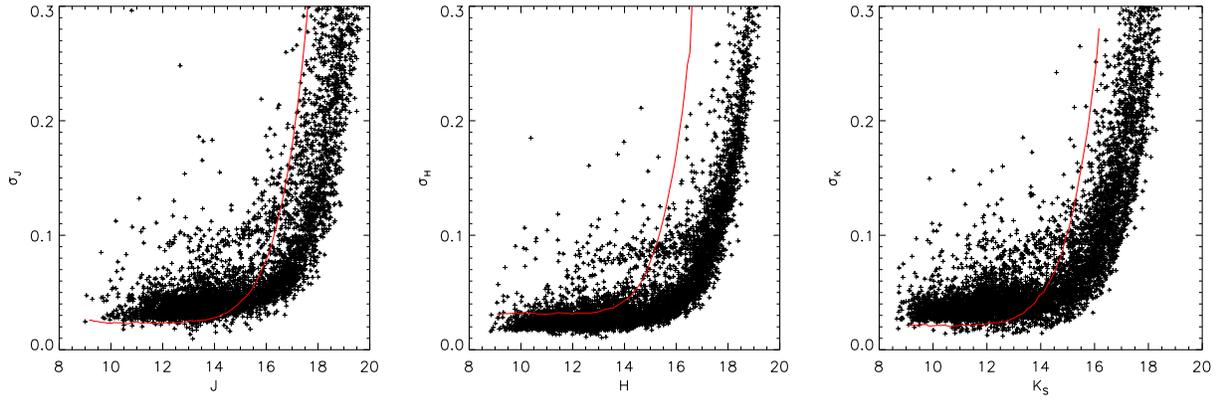}
 \caption{PSF photometric error as a function of magnitude for our point-like sources. The solid line indicates the locus of the average relation for the 2MASS\ sources falling in the ISPI\ field.}\label{fig:magerrcal}
\end{figure*}

 Figure \ref{fig:magerrcal} shows the photometric errors plotted as a function of magnitude for all point-like sources. The multiple threads in the distribution of the photometric errors are due either to the superposition of deep and short exposures (bright stars,  measured in short 3~s exposures, turn out to have relatively larger errors) or to the sources measured, with higher photometric errors, in the crowded central part of the cluster. For comparison we also plot the average mag vs. Delta mag relation for the 2MASS\  sources falling in our field.

\section{Completeness}\label{sec:compl}
To determine the completeness of our photometric catalog, we used an artificial star experiment. For each field and filter, we averaged the 9 PSFs created for photometric extraction (Section \ref{sec:datasec}) into a common reference PSF valid for the full image. We scaled the reference PSF in steps of 0.1~mag covering the entire magnitude range measured across each field and injected each scaled PSF\  at a random position in the image, with the only caveat of leaving enough distance from the border to allow for a meaningful measure. We assumed as a detection criterion a photometric error less than 0.3~mag
%, a value that we verified to be nicely compatible even with our early %visual inspection. We also required 
and less than 0.5~mag difference between the injected and recovered magnitude,   making also sure that the stars recovered by the code were the artificial ones and not previously known    real stars.  The completeness was calculated as the fraction of successful detection after 1000 iterations of the process, for each magnitude bin.

\begin{figure}
 %\epsscale{2}
\plotone{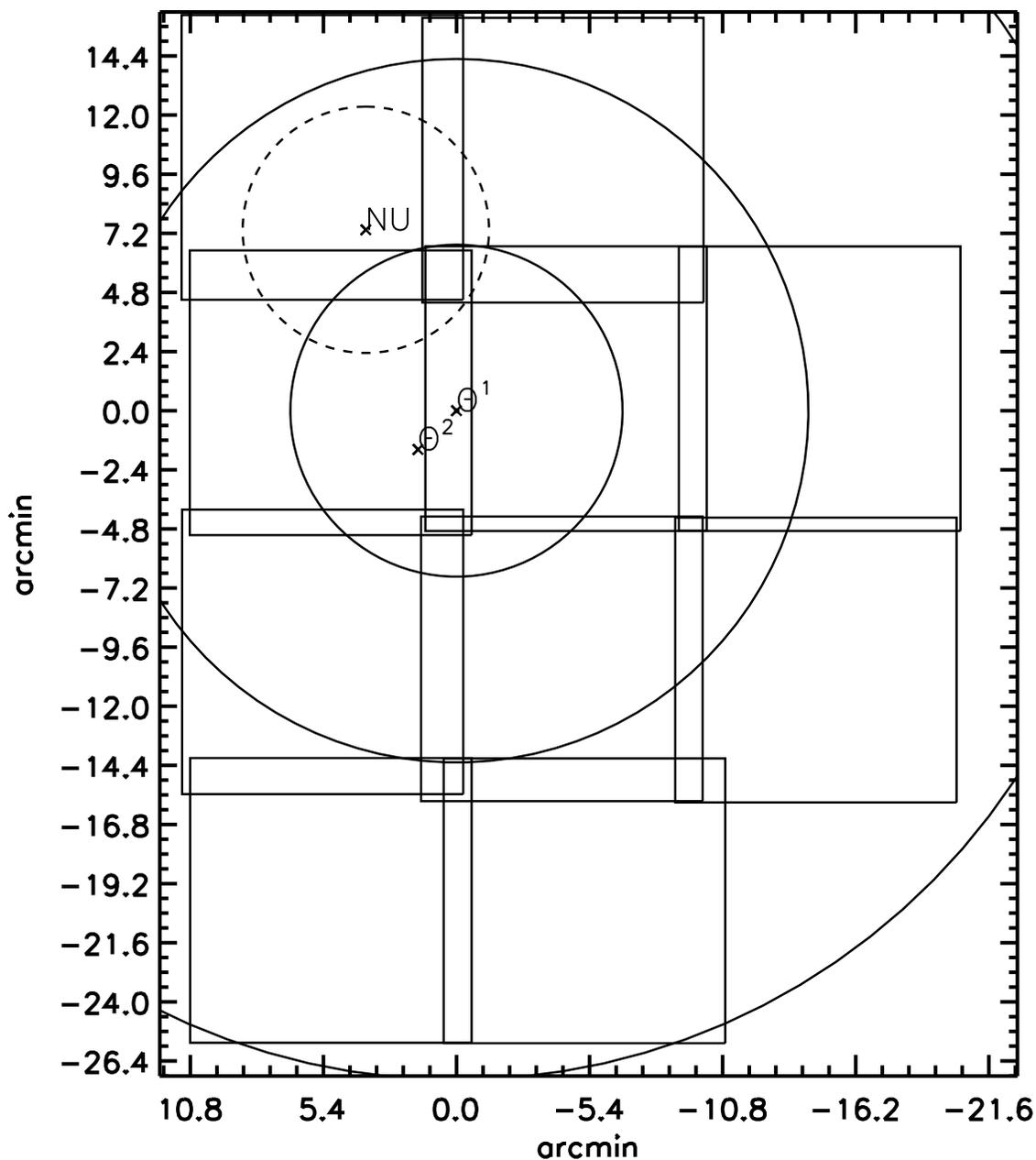}
 \caption{Projected regions identified in our surveyed area. The solid circles define the three radial areas containing an equal number of point-like sources, with radii of 6.7\arcmin, 14.3\arcmin\ and 27.2\arcmin, respectively. The dashed circle indicates the exclusion field centered around NU~Ori. The positions of $\theta^1$~Ori-C, $\theta^2$~Ori and NU~Ori are also indicated, together with our original field positions. The axes show the angular offsets from $\theta^1$~Ori-C (RA~(2000.0)=$05^h 35^m 16.46^s$, DEC~(2000.0)=$-05^\circ 23\arcmin 23\farcs18$).
}\label{fig:cyl}
\end{figure}

It turns out that the completeness (and the sensitivity)\ of our survey is significantly affected by the nebular brightness. We have therefore defined three concentric regions (Figure \ref{fig:cyl}) at increasing distance from $\theta^1$Ori-C. The limiting radii, respectively 6.7\arcmin, 14.3\arcmin\  and 27.2\arcmin\ (corresponding to projected distances of 0.81, 1.72 and 3.27~pc at a distance of 414~pc, \cite{Menten07}) have been set in such a way that the three regions  contain the same number of point-like sources. In order to avoid significant contamination from the M43 cluster, we neglect in our analysis a circle with a $5\arcmin$ radius centered on NU~Ori (RA(2000.0)=$05^h 35^m 31.37^s$, DEC(2000.0)=$-05^\circ 16\arcmin 02\farcs6$, see Figure~\ref{fig:cyl}). 
       
\begin{figure*}
 \epsscale{1}
 \plotone{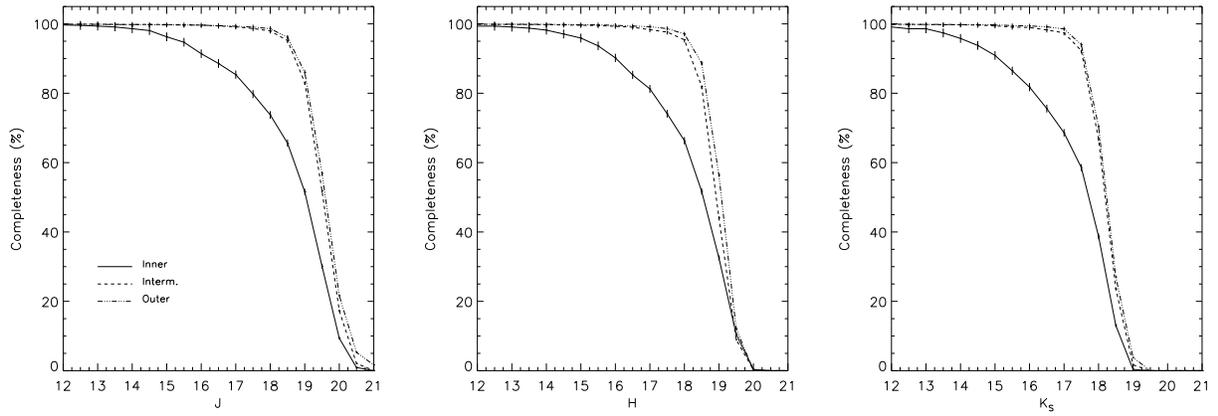}
 \caption{Completeness levels for the \jb~(left), \hb~(center) and \kb~(right) bands derived from the artificial star experiment. Each plot shows the results for the 3 radial regions.}\label{fig:complet}
\end{figure*}

\begin{figure*}
 \centering
 \includegraphics[width=.9\linewidth]{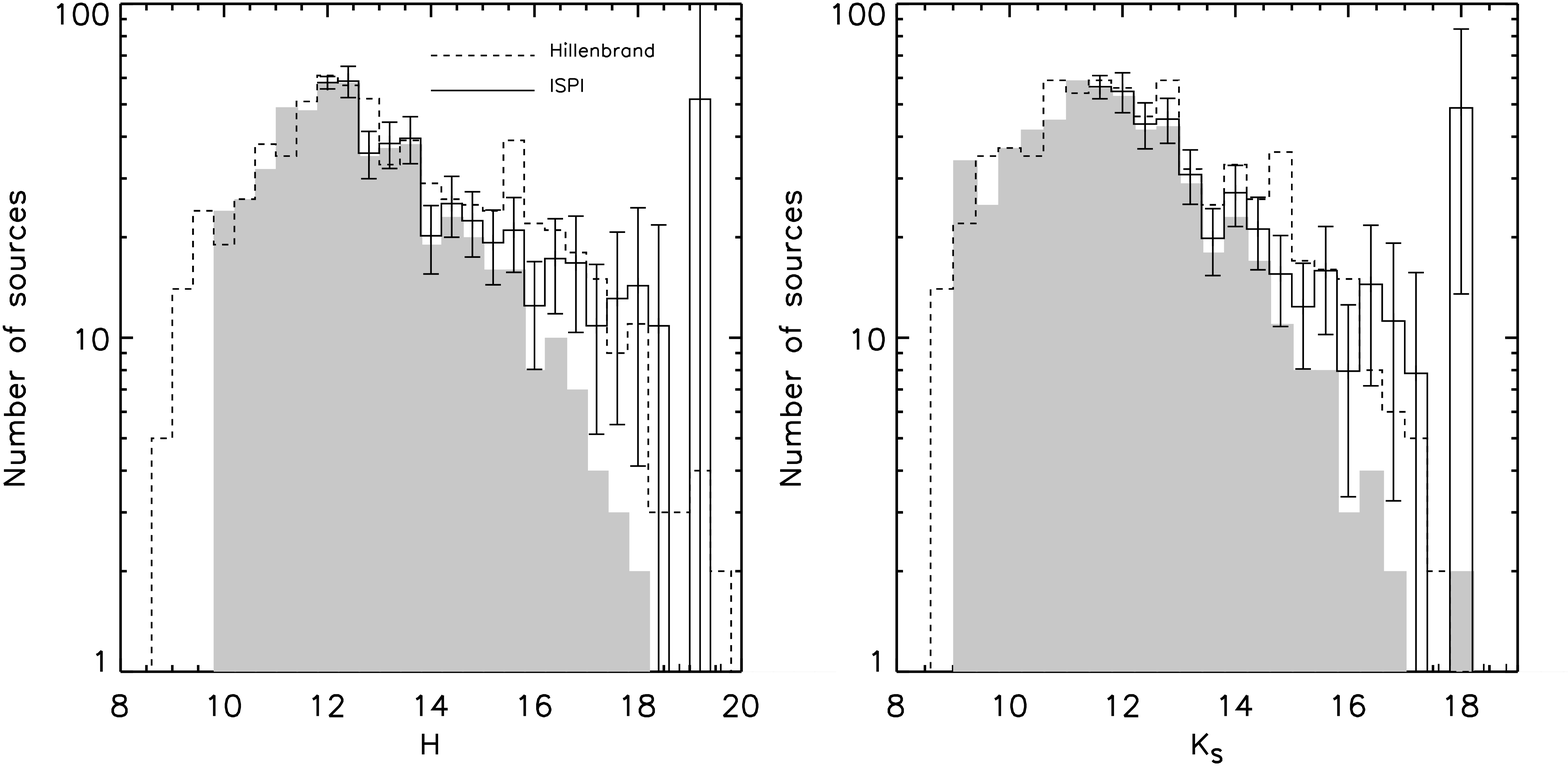}\\
 \includegraphics[width=.9\linewidth]{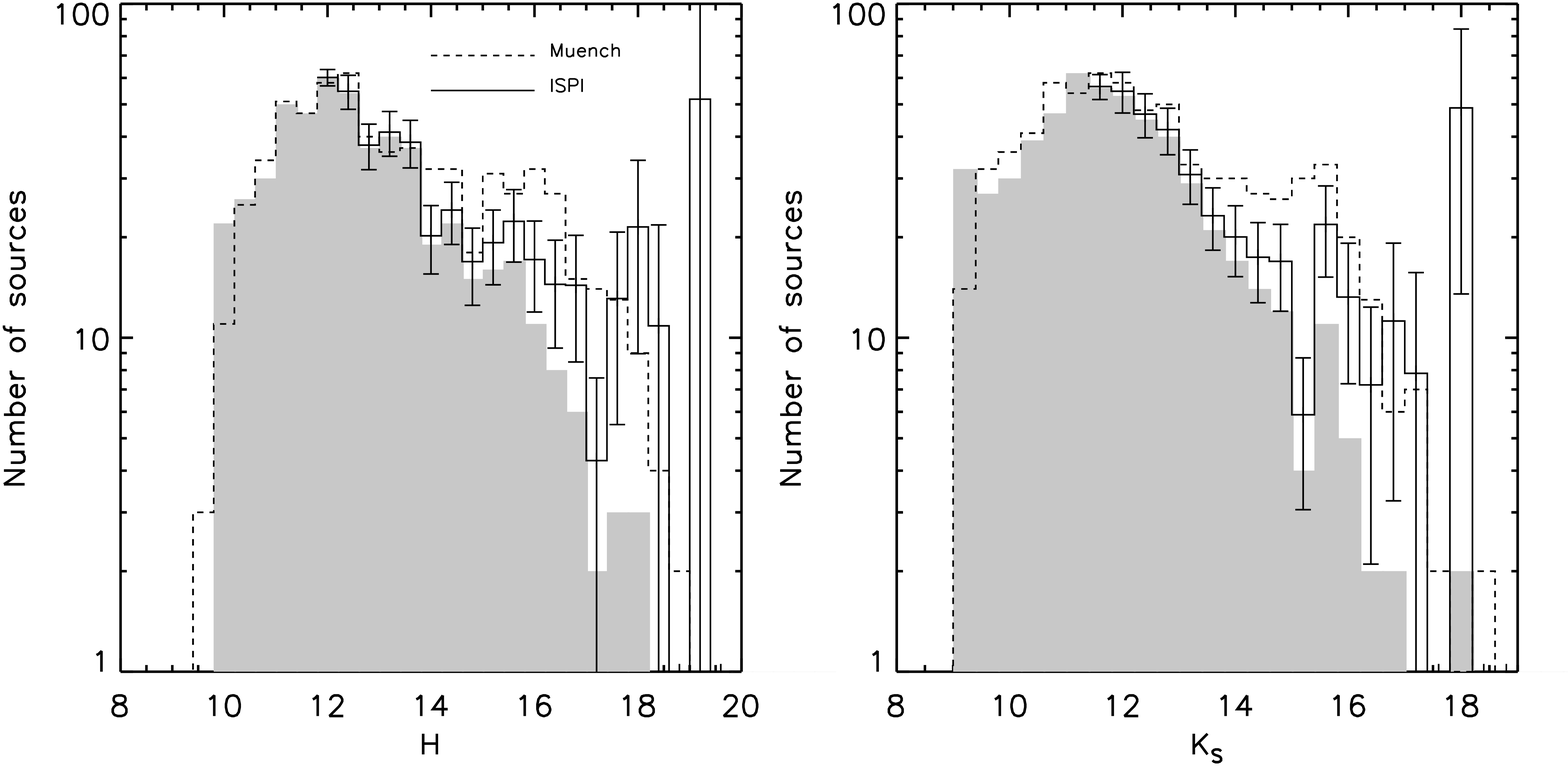}
 \caption{dashed lines:\ \hb\ and \kb\ luminosity functions for the \citet{HC00} (upper panels) and \citet{Mue02} (bottom panels) catalogs; gray area: luminosity functions for the same filter and fields from the ISPI\ observations (point sources only); solid line:\ ISPI\ luminosity functions corrected for completeness. The error bars have been obtained by adding quadratically the statistical (poissonian) uncertainty on the measured counts to the completeness correction error derived from the Monte Carlo simulations.}\label{fig:compare}
\end{figure*}

The results of our simulations  are shown in Figure \ref{fig:complet}. For each filter, the completeness estimated in the outer region is slightly, but systematically, larger than that in the intermediate region, which in turn is much larger than in the bright inner region, the most heavily affected by crowding and diffuse emission, where our sensitivity begins to drop at \hb\kb\ $\sim14$~mag.  A\ comparison with \cite{HC00} and \cite{Mue02}, who surveyed  the inner part of the cluster ($5\arcmin\times5\arcmin$ and 
$6\arcmin\times6\arcmin$ fields, respectively) deriving $HK$ photometry down to our photometric limit ($\kb\sim18$~mag), shows that we detect as point sources more than 90\% of their sources (648/706 in the \hb-band and 657/698 in the \kb-band for\ \cite{HC00} and 601/662(\hb) and 665/714(\kb) for \cite{Mue02}). Figure \ref{fig:compare}, where we overplot the Luminosity Functions (LFs) obtained by \cite{HC00} and \cite{Mue02} (dashed lines) to our observed LFs (gray area) constructed for their same regions, confirms that the missing sources are generally fainter than $\hb\kb\simeq14$~mag (the missing bright sources, detected but not measured by ISPI due to saturation, are not relevant here). If we apply to our \hb\ and \kb\ LFs the appropriate completeness corrections, estimated by repeating the artificial star  experiment on the same fields of  \cite{HC00} and \cite{Mue02}, our LFs\  become consistent with theirs, with the exception of the secondary peak at $\kb\simeq15.5$~mag discussed by \cite{Mue02}. In Section~\ref{sec:lf} we will apply the completeness corrections to the measured counts, when we will compare the LFs at various distances from the cluster center. 

%For fainter magnitudes, differences arises probably due to differences in the completeness estimation. For example, \citet{HC00} run their artificial star test avoiding fake stars to be within 0.5 times the FWHM of their typical PSF from every known star. This criterion leads them to state that a 90\% completeness level is reached at \kb$\sim17.7$ for a 7$\sigma$ detection threshold. This result is somewhat in contrast with our 10\% level at the same magnitude and in the same region, due to the fact that we randomly populated the whole area, accounting for a mis-detection when the fake star is blended from brighter stars. In any cases, even in the faint domain, the three surveys leads to consistent results, making us confident of the algorithm used to compute the completeness level of our survey.

\section{Results}\label{sec:res}

Our final photometric catalog contains 7759 sources. Having visually inspected all sources, we find that 6630 can be classified as point-like sources and 933 as diffuse objects. The other 174 sources, typically brighter than $K_S\simeq10$, turn out to be saturated  even in our short 3~s images. We included them in our catalog, adopting the photometry and coordinates reported by 2MASS.
We also found that the 2MASS photometric catalog  misses 
22 sources in the Trapezium region clearly visible in the 2MASS images and saturated even in 
our short exposures. We have added their \jb\hb\kb\ photometry using the \cite{Mue02}
photometry. 

In Table \ref{tab:catalog} we summarize the number of sources measured in different combination of filters, exposure times, and morphological types. 
The source catalog, presented in Table \ref{tab:samplecat}, is available in the electronic edition of this paper. In Appendix \ref{sec:database} we illustrate the content of each field of the source catalog.

\begin{figure}
%\epsscale{}
 \plotone{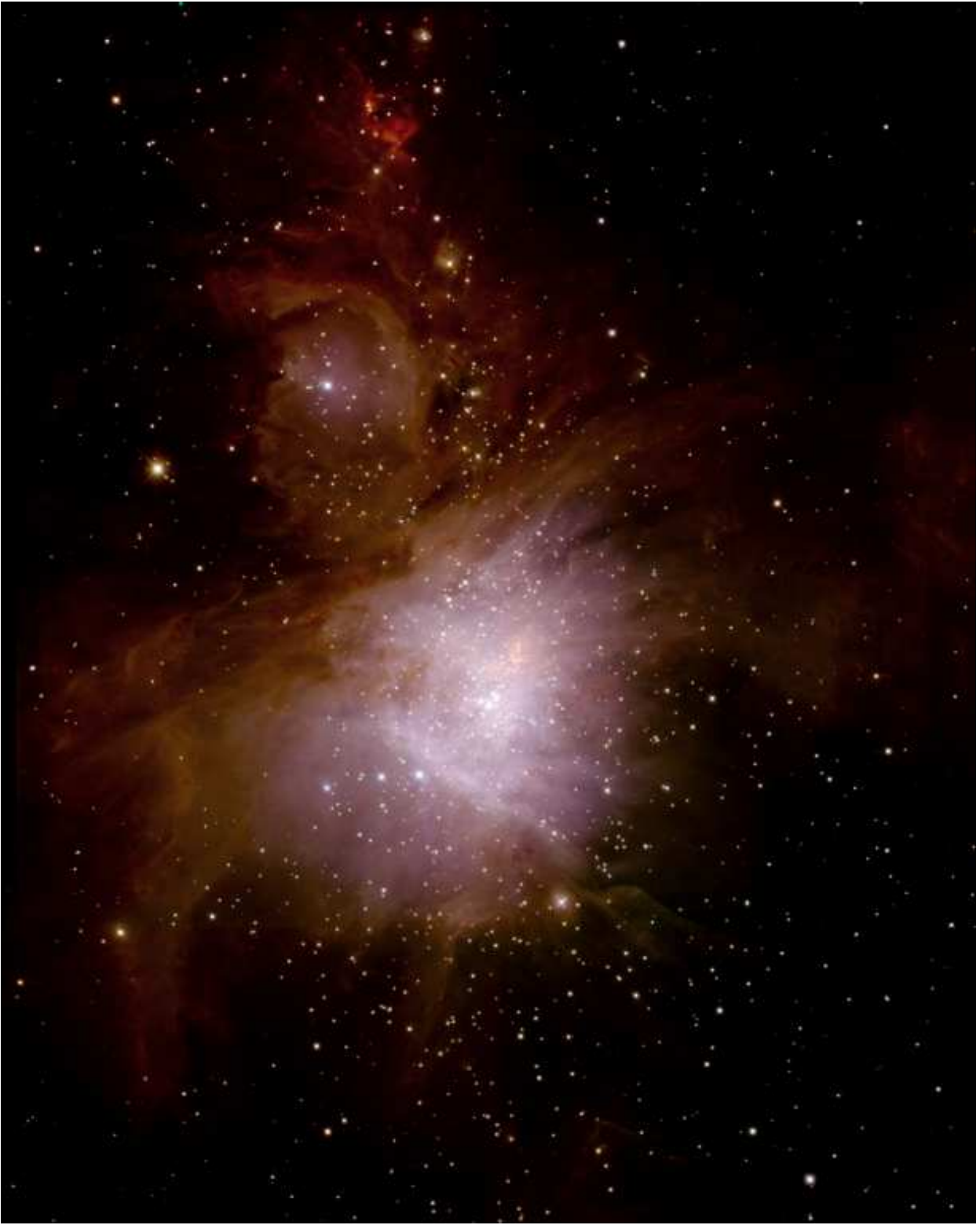}
 \caption{Near-IR color composite mosaic of the Orion Nebula from our ISPI\ images. The image has  been digitally enhanced to improve the visibility of the stellar sources in the central region. The RGB\ colors code the \kb, \hb\ and \jb\  bands, respectively. }\label{fig:orionrgb}
\end{figure}

\begin{figure}
%\epsscale{}
 \plotone{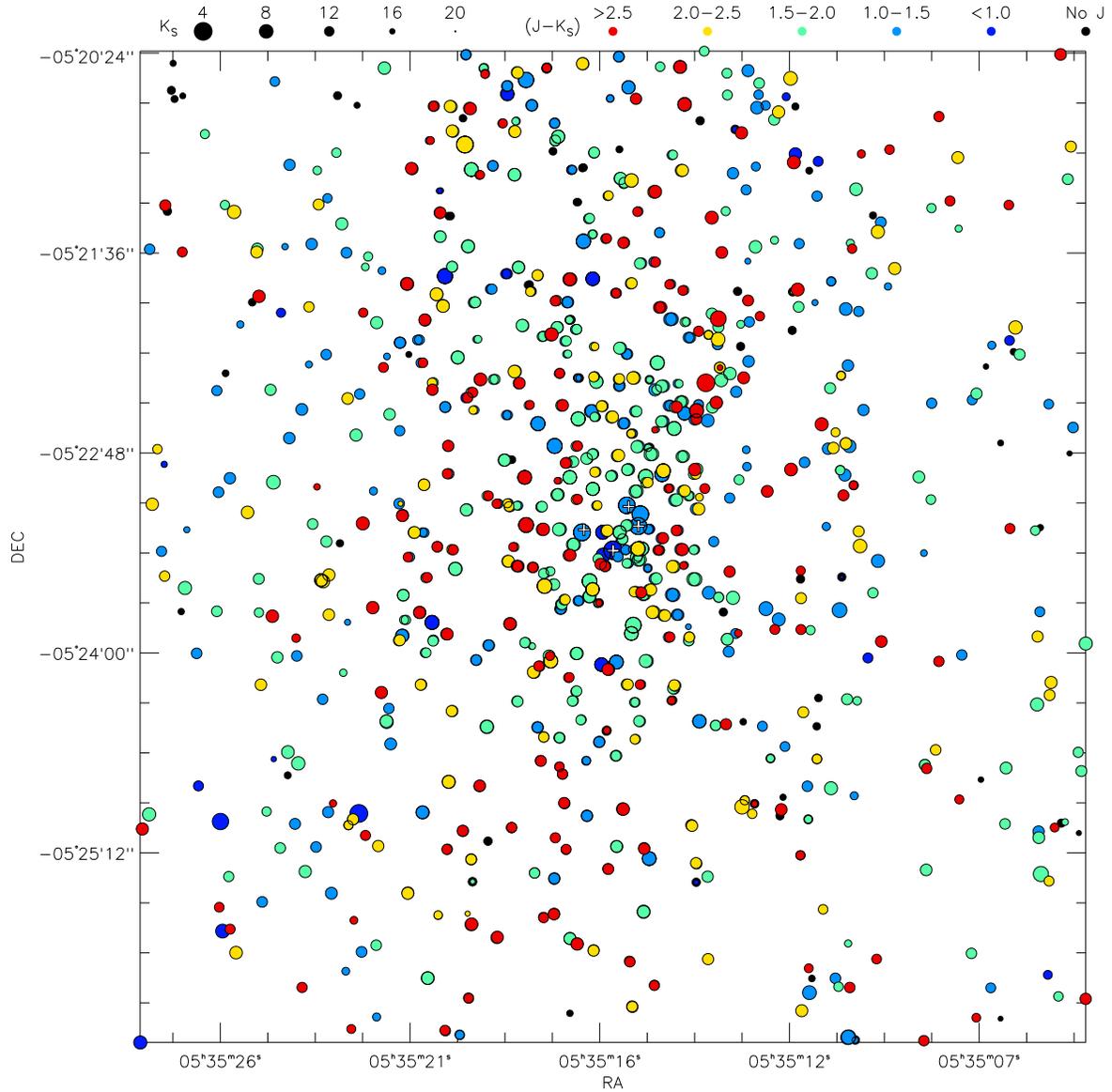}
 \caption{Spatial distribution of the ONC\ stars for the inner (Trapezium) region. The caption at top of the figure illustrates how the size of the circles is proportional to the source brightness, whereas the colors are from blue to red  according  to their observed \jb-\kb~ index. White arrows are used to indicate the Trapezium stars, for orientation}\label{fig:findingmap}
\end{figure}

 Figure~\ref{fig:orionrgb} shows a \jb\hb\kb\ color composite of our imaged field. The image, produced by the graphic staff  of the Space Telescope Science Institute, has been artificially enhanced to reduce the original dynamic range and more clearly reveal the inner part of the cluster. Figures~\ref{fig:findingmap}-19 provide 12 finding charts, starting from the Trapezium region (Figure~\ref{fig:findingmap}) and continuing with the 11 fields (Figure~9 to 19). The sources are coded with circles having sizes proportional to their brightness and colors scaled  from blue to red  according  to their observed \jb-\kb\ colors.

\section{Discussion}\label{sec:discussion}
%We now want to study the first results obtainable from our catalog. 

In this paper we present a preliminary analysis of the data, concentrating  on the point-like sources of the Orion Nebula Cluster having the full set of \jb\hb\kb\ photometry with magnitude errors $\Delta m \le0.3$, with the exception of sources seen only in Field 3 (thus lacking the deep \jb\ exposure) and  within $5^\prime$ from NU~Ori, i.e., lying in the field of the satellite nebula M43, about $15^\prime$ northeast of the Trapezium. Our sub-sample is therefore composed by 4103 sources.
\subsection{Color-color diagrams}\label{sec:ccd}

In Figure\ \ref{fig:ccd}a we show the near-IR\ {\sl \jb-\hb} versus {\sl \hb-\kb} color-color diagram. Following \citet{Lee05}, we overplot the intrinsic colors of the main-sequence stars and giant branch, together with  the locus of classical T~Tauri stars (CTTSs) surrounded by circumstellar disks   \citep{Meyer97}, given by the relation {$(\jb-\hb)-0.630(\hb-\kb)-0.497=0$} in the 2MASS photometric system. We also plot the interstellar reddening vector of  \citet{Cardelli89}, given by the relation $E(\jb-\hb)/E(\hb-\kb)=1.83$ in the 2MASS\ photometric system. Accounting for extinction, the locus of the CTTSs lies between the two parallel dotted lines defined  by the equations $(\jb-\hb)-1.83(\hb-\kb)+0.098=0$ and $(\jb-\hb)-1.83(\hb-\kb)+0.50=0$.

%, where \jb, \hb, and \kb\ are in this case 2MASS magnitudes, and the slope is specified by the interstellar reddening law given by \citet{Rie85}.

%Thus, the dashed parallel lines separate CTTSs from Herbig Ae/Be and from reddened main-sequence stars. \citet{Lee05} have shown that different kinds of PMS stars occupy separate regions in the \jb-\hb\ versus \hb-\kb\ CC diagram, enabling to identify candidate young stars in star-forming regions. LA SELEZIONE NON E' L'OBBIETTIVO DI QUESTO PAPER.

%)

For 45.2\% (1853/4103) 
of the sources, the near-IR\ colors are consistent with those of stellar dwarfs (luminosity class V), giants\ (luminosity class III) or CTTSs reddened by various amounts of foreground extinction. For 11.4\% 
(466/4103) 
of the sources the  near-IR\ excess is compatible only with the reddened  CTTSs, whereas only 0.8\%
 (35/4103)
 of the sources show near-IR excess significantly greater than those of CTTSs, indicative of strong \kb-band excess.

%The observed colors of these objects, substantially redder than the expectations from pre-main-sequence isochrones, can be attributed to a combination of extinction and near-infrared excess due to a circumstellar disk \textbf{, as discussed in Section X DOVE QUANDO PERCHE'}.

To assess how the observed reddening and color excess  vary with the distance from the center of the cluster, 
%assumed to be coincident with the position of $\theta^1$Ori-C (RA~(2000.0)=$05^h %35^m16.46^s$, DEC~(2000.0)=$-05^{\circ} 23^\prime 23\farcs18$), 
we consider the Hess diagrams for the three radial regions. The inner field (Figure~\ref{fig:ccd}b) shows a peak well detached from the main sequence, with  iso-density contours elongated in a direction parallel to the CTTSs locus, together with a long tail of stars with high reddening.   The Hess diagram for the intermediate region (Figure\ \ref{fig:ccd}c)  shows a main peak much  closer to the locus of  dwarf stars, together with a secondary  maximum at higher extinction. The intermediate contour levels (e.g. the green area) are now elongated  along the range of reddened main sequence stars. 
%reaching   reddening values as high as $A_V\simeq15$.
Finally, the Hess diagram for the outer region (Figure~\ref{fig:ccd}d)   shows again a strong peak compatible with dwarf stars affected by a small amount of extinction, whereas the low contour levels appear less pronounced than in the two other regions and  well within the limits of reddened main sequence or giant stars.

Overall, the three color-color diagrams can be interpreted as follows:\ 1)\ the inner Trapezium region is mostly populated by young stars with near-infrared excesses typical of CTTSs. They have generally  relatively modest amounts of extinction, which is compatible with the fact that the ionizing radiation from the Trapezium stars  has cleared the molecular cloud and exposed them to our view;  2)\  the intermediate region contains  a relatively larger number of heavily reddened sources. They  are seen through denser parts of the OMC, not yet reached by the expansion of the ionized cavity; 3) the outer regions mostly contain background objects seen trough the low opacity layers   at the edges of the main OMC ridge, which crosses the ONC approximately in north-south direction \citep{Goldsmith+97}.\

\subsection{Color-magnitude diagrams}\label{sec:cmd}
In Figure\ \ref{fig:hr} we present the (\hb, \jb-\hb) and (\kb, \hb-\kb) color-magnitude diagrams  still for the same sample. Together with our data, we plot the 1~Myr isochrone of \citet[{\tt CBA00}]{CBAH00}, which covers the range $0.001~M_\odot<M<0.1~M_\odot$ and is tailored to model substellar and planetary mass objects in the limit of high opacity photospheres. We also plot the 1~Myr isochrone of \citet[{\tt sie00}]{Sie00} ranging from $0.1~M_\odot$ to $7~M_\odot$, well-suited for stellar sources.
Both isochrones have been converted to the 2MASS photometric system using the transformations of \citet{Carp01}. The offset between the two isochroness can be regarded as a graphic representation of the uncertainties in theoretical models. An $A_V=10$ reddening vector is also indicated, as well as the reddening lines starting from the isochrones in correspondence of a $1~M_\odot$  object (\jb=10.50, \hb=9.88 and \kb=9.76 at 414~pc in the 1Myr \texttt{Sie00} isochrone), a M=0.075~$M_\odot$ object at the hydrogen-burning limit mass (\jb=14.29, \hb=13.86 and \kb=13.56 at 414~pc in the 1Myr \texttt{CBAH00} isochrone) and a $M=0.012~M_\odot$  substellar object at the  deuterium-burning limit mass (\jb=17.33, \hb=16.90 and \kb=16.48 at 414~pc in the 1~Myr \texttt{CBAH00 }isochrone). For this distance and isochrone, we are therefore sensitive to stellar sources down to $A_V\simeq 60$, brown dwarfs down to $A_V\simeq40$, and planetary mass objects down to $A_V\simeq 15$.

%, in opposition to their \texttt{COND} model, which does not take into account opacities of dust grains in the photosphere. The \dusty\ and \texttt{COND} models represent extreme situations which bracket the more likely intermediate case resulting from complex thermochemical and dynamical processes.

Almost all data points lie to the right of the isochrones and are therefore compatible with reddened 1~Myr old sources (as well as heavily reddened field objects).
If we discard the blue sources with $\jb-\hb<0.3$, typically affected by large measurement errors, we find that the reddening lines identify 2246 sources 
%(2603 sources if we include the M43 region)  
in the region of reddened  stellar photospheres, 1298 
%(1533) 
sources in the region of reddened brown dwarfs and 142 
%(188) 
sources in the region of reddened planetary mass objects.
In the (\kb, \hb-\kb) color-magnitude diagrams these values are
2134,
%(2474), 
1145, 
%(1349)\ 
and 421,
%(517)
respectively.

%173,1357,2212(J,J-H)  483,1199,2074(K,H-K) nei due diagrammi, senza m43
%188,1533,2603(J,J-H)  517,1349,2474(K,H-K) nei due diagrammi, con m43

Splitting the cluster in the same three regions of Figure~\ref{fig:cyl},
% \ref{sec:ccd} 
we obtain the color-magnitude diagrams presented in Figure~\ref{fig:cmhess}. From top to bottom, Figure~\ref{fig:cmhess} shows the (\hb, \jb-\hb) (left column) and (\kb, \hb-\kb) color-magnitude diagrams for the inner, median and outer regions, respectively. Consistent with what  was found analyzing the color-color diagrams, the inner and outer regions appear dominated by two different populations.\ The top diagrams clearly reveal the location of  the ONC: an ensemble of stellar sources with modest amounts of foreground extinction. The bottom diagrams are dominated by fainter sources (of course, for background stars the 1~Myr isochrone at 414 pc does not apply), whereas the intermediate region contains a mix of both populations. A careful look at the diagrams show that the peak corresponding to the ONC drifts down (i.e., to lower masses) and left (to lower extinction values)  moving from the inner to the outer regions. The shift to lower masses represents an overabundance of  higher mass sources at the center of the cluster, in agreement with the evidence for mass segregation found by \cite{HH98}. The redder color of  the cluster members in the central field may be  due to the fact that they are, on average, either more embedded in the OMC or subject to  higher circumstellar extinction from their heavily warped or flared circumstellar disks \citep{RBP02}, or  a combination of the two. 
%In any case, these effects  seem to  point to the central cluster as the region where star formation is more recent (or even still ongoing, e.g.  the sources in BN and Orion South). One could therefore coherently deduce a scenario in which   low mass stars started to form first in the outer regions, whereas star formation in the inner part followed somewhat later with a relatively higher fraction of massive stars. Such an evolutionary sequence of events could indicate that  mass segregation is primordial instead of temporary due to the random motions of massive stars, as recently suggested by \cite{Xin-yue+09}.

\subsection{Luminosity Functions}\label{sec:lf}

 We conclude our analysis showing the \jb, \hb~ and \kb-band luminosity functions (LFs) for all point-like sources. Once again, we remove stars less than $5^\prime$ projected distance from NU~Ori and divide the remaining sample in three groups according to the distance criteria adopted in the previous sections. Rather than assuming a certain completeness limit, we fully apply our completeness corrections, derived in Section~\ref{sec:compl}, to the source counts measured in the various regions.

%Our results  (Figure\ \ref{fig:lfcomplfig}) are generally consistent with the previous near-infrared studies of the ONC, with 

The histograms for  the inner region (Figure\ \ref{fig:lfcomplfig}, thick solid line) show a broad peak at $\jb\simeq13$,  $\hb \simeq 12$~ and $\kb\simeq 12$, corresponding to an M6 star (0.175~M$_\odot$) at 1~Myr, according to  the \cite{Sie00} model with zero reddening. The peak shows an abrupt drop at about $\jb\hb\simeq14$ and $\kb\simeq13$, approximately by a factor $\simeq2$, which is fully consistent with the previous near-infrared studies of the ONC, e.g. Figure~5 of \cite{HC00}. On the other hand, the luminosity functions corrected for completeness  remains remarkably flat across the substellar region down to our sensitivity limit. This is not what has been found by other authors who concentrated on the innermost area. 
%This is especially evident in the \kb-band,  whereas in the case of the \jb- and \hb-bands the  LFs in the last couple of bins. 
We remark that our completeness correction has the largest effect  in  the inner region,   the one more affected by crowding and by the brightness of the nebular background. On the other hand, we have seen that our completeness correction fails to reproduce the $\kb\simeq14$ secondary peak of \cite{Mue02}, and for this reason can hardly be suspected of providing an artificial overcorrection in the central area. 

Moving from the inner to the outer parts of our survey area, the main peak in the LFs becomes less prominent in the median region and eventually disappears in the outer regions giving to the LFs\ an increasing monotonic slope. If we compare our \kb-band LF\ for the outer cluster with the $K$-band LF\ of the off-cluster control field presented by \cite{Mue02} (Figure 3a of their paper), which shows no   stars brighter than $K\simeq12$, we see that our outer field still contains a fraction of sources belonging to the cluster or, more in general, associated to the recent star formation events in the Orion region (i.e. possible foreground members of the Orion OB1 association). This is in agreement with what we found in the previous section discussing the color-magnitude diagrams,  not by chance since the LF represents the projection on the vertical axis of the color--magnitude plot.  

%Turning our attention to the shape of the LF\ of the central region in the substellar regime, we see that the peak
%{, however, appears narrower in the \jb-band than in the \hb- and \kb-bands,and} shows an abrupt drop at about $\jb\hb\kb\simeq14$, approximately by a factor $\simeq2$. Again, this has a clear correspondence  in the color-magnitude diagrams (Figure~\ref{fig:cmhess}, top panels): at about 0.075$M_\odot$, the source density of the main cluster population suddenly drops, as evidenced by the change in color from green/cyan to blue.    After this drop, the LF\ decline is rather moderate  until the final fall approaching the sensitivity limit of the survey. In what concerns the outer fields, the LF\ shows either a double peak (intermediate region) or a monotonic increase to the faintest magnitudes (outer region), due to the increasing influence of sources not associated to the cluster.

   The flattening of the substellar LFs in the inner region may be due to field star contamination. To address this possibility, we  can use our outer field as a control field, knowing that it will probably provide an overestimate of the contribution of non-cluster sources and therefore, after subtraction, a lower limit to the cluster LF. This for two main reasons:\ first,  we have seen that our outer field still contains a fraction of cluster-related  sources and, second, background sources should be more easily observed in the outer field rather than in the central region due to the larger background extinction caused by the underlying OMC-2, which has the ridge of highest density  crossing the cluster core nearly in a north-south direction (see e.g. the extinction map obtained from the C$^{18}$O column density data of \citet{Goldsmith+97} presented in Fig 2 in \cite{HC00}). Figure~\ref{fig:lfcomplfig} (bottom, gray area) shows the LFs corrected in this way, having properly normalized  the source densities to the same surface area. There is a clear reduction of source density in the substellar regime and the main peak becomes more prominent, but the substellar LFs seem to remain flat or show, at best, a very modest negative slope. 

Alternatively, it is possible to account for the higher background extinction beyond the central field, and therefore remove at least some of the systematic bias introduced by using the outer field, by shifting the outer LFs, for each filter, by magnitude values compatible with the interstellar reddening law of \cite{Cardelli89}. This because with a proper, wavelength dependent, change of zero points one can use the outer field counts to reproduce  the luminosity functions that would have been observed with higher foreground extinction. If we  refer again to the \citet{HC00} extinction map, we see  that the very central part of the OMC-2 is characterized by a visual extinction as high as $A_V=75^m$. For a detailed analysis one should account for the spatial variation of the extinction, as done e.g. by \cite{HC00}, but for the moment let's simply assume the LFs\ of the outer region and put all sources under  $A_J\simeq 10$~mag,  $A_H\simeq 6$~mag, and $A_{K_S}\simeq4$~mag, corresponding to   $A_V\simeq33$~mag. It is immediate to see that shifting the monotonically increasing field LFs to the right by a few magnitudes and subtracting them  from the central ones produces no visible change on the central ones.
One can therefore safely conclude that the true LF of the inner region lies between the one derived above subtracting the outer field source density (corresponding to a fully transparent background) and the one originally observed (corresponding to a fully opaque background), which means that  the inner LFs overall remain quite flat in the brown-dwarfs regime (0.075-0.012$M_\odot$). In a future paper we will address the implications of this result on the cluster Inititial Mass Function, evaluating the foreground extinction and the stellar parameters of the point sources on the basis of the rich set of multicolor photometry obtained for the HST\  Treasury Program on the Orion Nebula.
%A full analysis, accounting for the background contamination due to both galactic and extragalactic sources and a more refined model for the OMC-2\ extinction, will be presented in a future paper (Scandariato et al., 2009, in preparation).

{
%More exhaustive analysis (e.\ g.\ by correcting for contaminants in the sample), it would supports the evidence for some mass segregation, i.\ e.\ the settling of massive stars towards the center of the cluster, in agreement with what \citet{Hill97} suggests and in disagreement with the finding of \citet{LR05} studying there deep survey in a region out of the Trapezium cluster, as well as with \citet{Gra,Grb,Grc}, which state that mass segregation is observable only for clusters older than $\sim$10Myr.

%Another possible explanation for the radial dependence requires massive stars to form more easily ``in situ'', as a result of a star formation efficiency (SFE) which depends on the pressure of the pre-stellar environment. As suggested by \citet{Lada08}, the critical mass for star formation decreases with the internal pressure of the cores. Therefore, massive star formation occurs more easily in higher pressure regions, i.e. at the center of nearly isothermal clouds.
}
%, against the formation of lower mass stars and brown dwarfs.
%where pressure makes scale radii of bound systems larger than in lower regions %where infalling material could be less bounded, and then escape more easily %from the gravitational field of the central object (MASSIMO: DEVO\ VEDERE\ %l'EQUAZIONE). 

\section{Summary and conclusions }
 We have presented a photometric survey of the Orion Nebula Cluster in the \jb-, \hb- and \kb-passbands carried out at the 4m telescope of Cerro Tololo. The survey, covering a field of about $30\arcmin\times40\arcmin$ centered about $1\arcmin$\ southwest of the Trapezium, has been performed in parallel to visible observations of the same region made in La Silla (Da Rio et al. 2009). The two datasets constitute the first panchromatic survey covering simultaneously the Orion Nebula Cluster from the {\sl U}-band to the \kb-band. \

The final catalog, photometrically and astrometrically calibrated to the 2MASS\  system, contains  7759 sources, (including 174 and 22 sources whose photometry has been taken directly from the 2MASS\ and \cite{Mue02} catalogs, respectively). This represents the largest near-IR\ catalog of the Orion  Nebula Cluster to date. Our sensitivity limits allow to detect  objects of a few Jupiter masses under about $A_V\simeq 10$ magnitudes of extinction.

We present the color-color diagrams, color-magnitude diagrams and the  luminosity functions for three regions centered on the Trapezium containing the same number of sources (excluding the M43 sub-cluster). Sources in the inner region typically show IR\ colors compatible with reddened T\ Tauri stars, whereas the outer fields are dominated by field stars seen through an amount of extinction  which decreases with the projected distance from the  center.
The color-magnitude diagrams allow to clearly distinguish between the main ONC population, spread across the full field,  and background sources. 
%The position of the ONC\ peak slightly drifts to lower mass and bluer colors  moving from the inner to the outer region, suggesting that the inner cluster contains a higher fraction of massive and young stars. This  points to a scenario in which star formation in the ONC has proceeded from the outside to the inside increasing the efficiency of massive star  formation. 
After correction for completeness, the luminosity functions in the inner region remains nearly flat, with marginal contamination from background sources. 

\acknowledgments
We are grateful to the ESO\ director for allocating Director Discretionary Time with the ESO-MPIA\ 2.2m telescope in parallel with the ISPI\ time at Cerro Tololo. We are grateful to the ESO\ staff at La Silla for carrying out the WFI\ observations in service observing mode and to N. van der Bliek and the CTIO staff for support with the observations and data calibration. Observing time with ISPI was obtained within the framework of the HST Treasury Program GO10246.

The reduced images presented in this paper can be downloaded 
from the  High-Level Science Products archive hosted at the Multimission Archive at STScI at the address

http://archive.stsci.edu/prepds/orion/

\appendix\  

\section{The database}\label{sec:database}

The following is a legend of the fields in the database. 

\begin{enumerate}
\item{\sl id}: Running catalog number. The 174 sources taken from the 2MASS\ catalog  have a progressive {\sl id}  starting from 10,000. Sources from the \cite{Mue02}\ catalog have their original entry number added to either 20,000 or 21,000, depending if the photometry was derived from FLWO\ or VLT\ observations (for sources having both measurements, the FLWO\  photometry has been adopted).
%Note that these values begin at 1 and not 0, and that any database routines %will use entry numbers to refer to stars.
\item {\sl ra}: Source Right Ascension (2000.0)\ in decimal hours. Based on the \kb\ band value if the source is detected and not saturated,  otherwise the \hb-band or the \jb-band, in the order.
\item {\sl dec}: Source Declination (2000.0) obtained following the same priorities of the RA above.
\item {\sl raj}: Right Ascension in the \jb\ band images.
\item {\sl decj}: Declination in the \jb\ band images.
\item {\sl rah}: Right Ascension in the \hb\ band images.
\item {\sl dech}: Declination in the \hb\ band images.
\item {\sl rak}: Right Ascension in the \kb\ band images.
\item {\sl deck}: Declination in the \kb\ band images.
\item {\sl j}: \jb\ band PSF magnitude, calibrated to the photometric 2MASS system.
\item {\sl dj}: Uncertainty in the above.
\item {\sl h}: \hb\ band PSF magnitude, calibrated to the photometric 2MASS system.
\item {\sl dh}: Uncertainty in the above.
\item {\sl k}: \kb\ band PSF magnitude, calibrated to the photometric 2MASS system.
\item {\sl dk}: Uncertainty in the above.
\item {\sl j\_ap}: \jb\ band magnitude for a 5 pixel radius
 aperture, calibrated to the photometric 2MASS system.
\item {\sl dj\_ap}: Uncertainty in the above.
\item {\sl h\_ap}: \hb\ band magnitude for a 5 pixel radius aperture, calibrated to the photometric 2MASS system.
\item {\sl dh\_ap}: Uncertainty in the above.
\item {\sl k\_ap}: \kb\ band magnitude for a 5 pixel radius
 aperture, calibrated to the photometric 2MASS system.
\item {\sl dk\_ap}: Uncertainty in the above.
\item {\sl image}: Image from which the measures are taken. Image numbers run from 1 to 11.
\item {\sl exptime}: Set to either 3 or 30, it refers to the exposure time in seconds of the individual frames building the image from which the photometry has 
been taken. Sources with exptime=3 are generally saturated in the 30~s exposures.
\item {\sl status}: Flag sources according to various criteria:
        \begin{itemize}
        \item[$\rightarrow$] { 1 }: source is point-like, detected during the first photometric extraction;
        \item[$\rightarrow$] { 2 }: source is point-like, detected after PSF subtraction;
        \item[$\rightarrow$] { 3 }: source appears extened (galaxies, proplyds, diffuse objects in general);
        \end{itemize}
% Objects which have been examined and are believed to be real have status=0.
%   Objects with status=1 are still to be examined.    Objects believed to be real but are
%   duplicates in the border regions of other sources already included
%   from a different field section have status=2. Objects with status
%   greater than 2 are various types of artifacts and should be ignored.
 
%I wrote a program to display a given star, called
%viewstar.pro.  It simply takes a database ID number as an argument.
%That program has an issue with the full color drawing but that is
%because each image is scaled when displayed and each filter is not
%treated equally.
\item {\sl clone}: Flag for multiple detections. Sources detected multiple times have the {\sl clone} value equal to the {\sl id} value  of their first appearance in the database, multiplied by -1.
\end{enumerate}

{\it Facilities:} \facility{CTIO (ISPI)}

\clearpage

%%%
%FIGURE\ 1
%
\clearpage

%%%
%FIGURE\ 2
%
 \clearpage

%%%
%FIGURE\ 3
%
\clearpage

%%%
%FIGURE\ 4
%
\clearpage

%%%
%FIGURE\ 5
%
\clearpage

%%%
%FIGURE\ 6
%
\clearpage

%%%
%FIGURE\ 7
%
\clearpage

%%%
%FIGURE\ 8
%
\clearpage

%%%
%FIGURE\ 9
%
\begin{figure}
%\epsscale{}
 \plotone{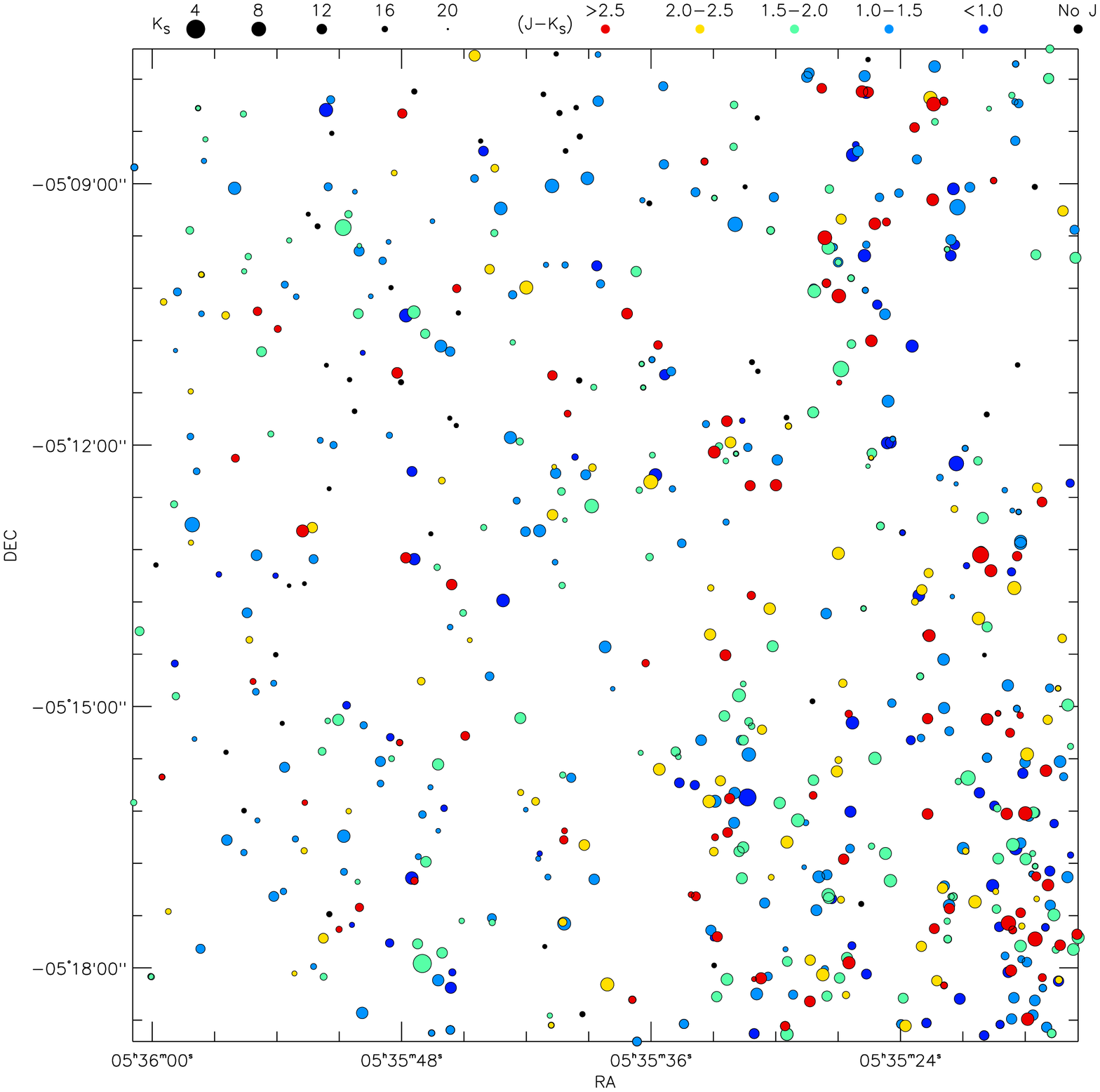}
 \caption{Same as Figure~\ref{fig:findingmap}, for Field 1.}
\end{figure}
\clearpage

%%%
%FIGURE\ 10
%
\begin{figure}
%\epsscale{}
 \plotone{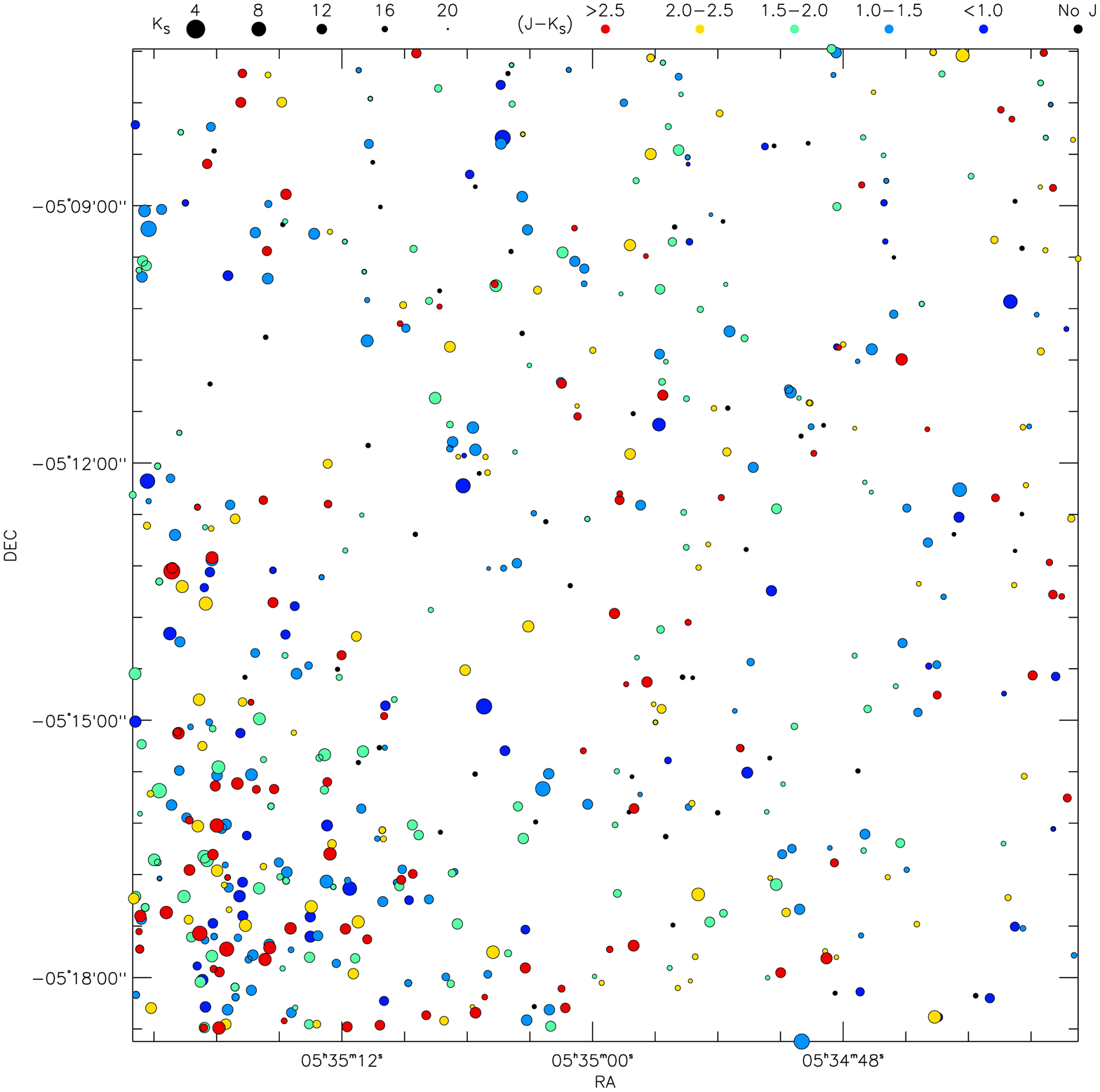}
 \caption{Same as Figure~\ref{fig:findingmap}, for Field 2.}
\end{figure}
\clearpage

%%%
%FIGURE\ 11
%
\begin{figure}
%\epsscale{}
 \plotone{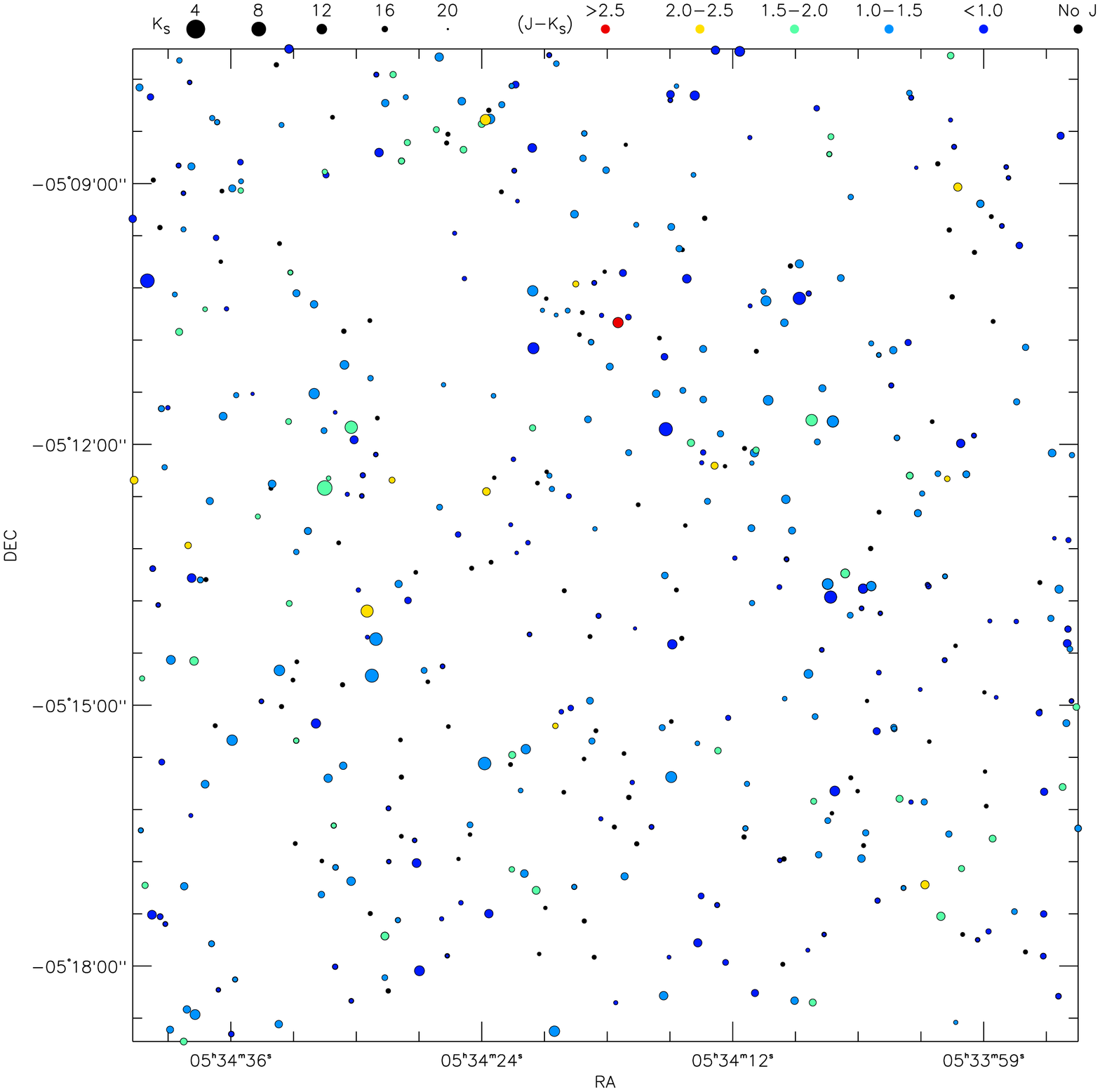}
 \caption{Same as Figure~\ref{fig:findingmap}, for Field 3.}
\end{figure}
\clearpage

%%%
%FIGURE\ 12
%
\begin{figure}
%\epsscale{}
 \plotone{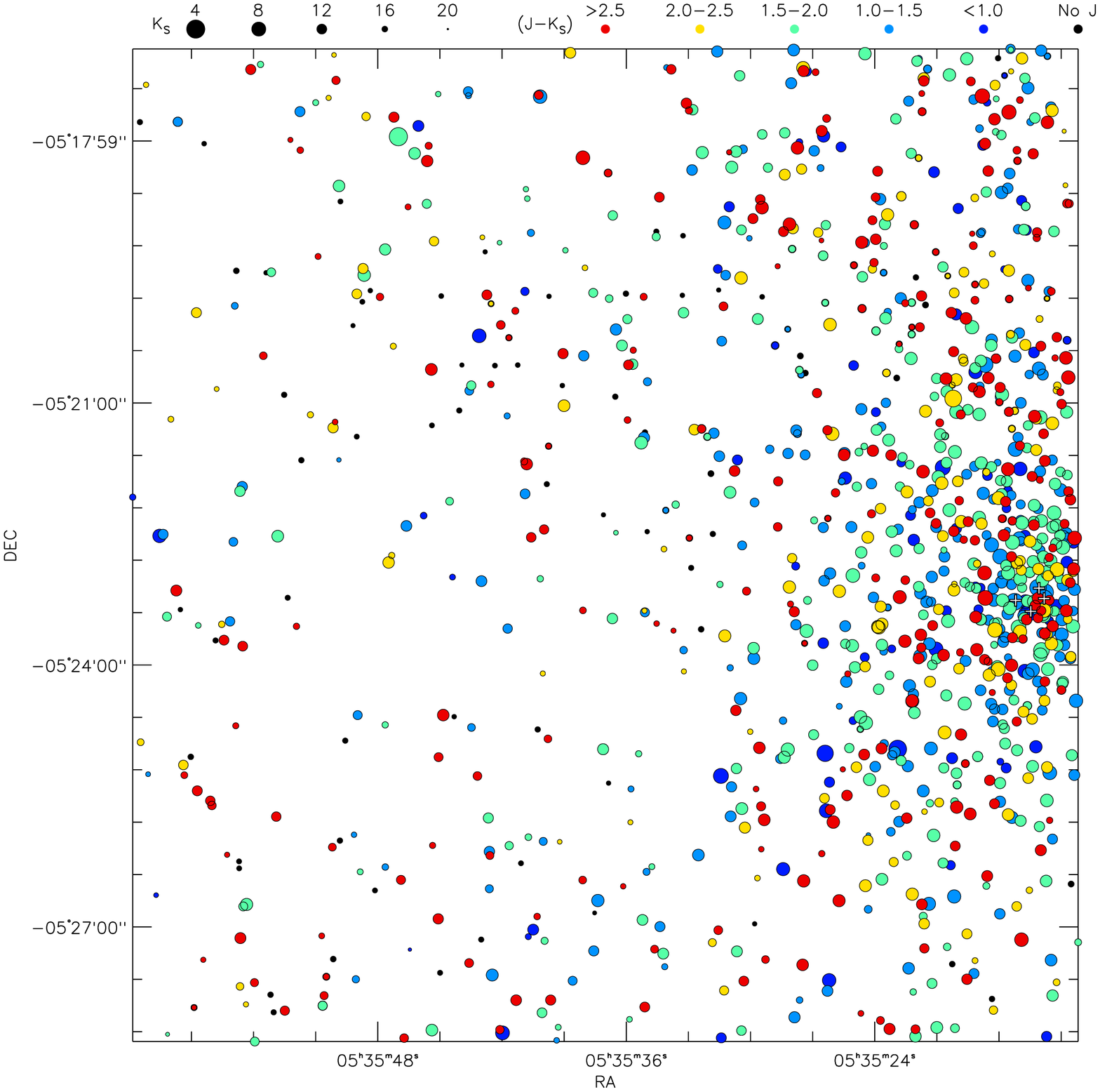}
 \caption{Same as Figure~\ref{fig:findingmap}, for Field 4.}
\end{figure}
\clearpage

%%%
%FIGURE\ 13
%
\begin{figure}
%\epsscale{}
 \plotone{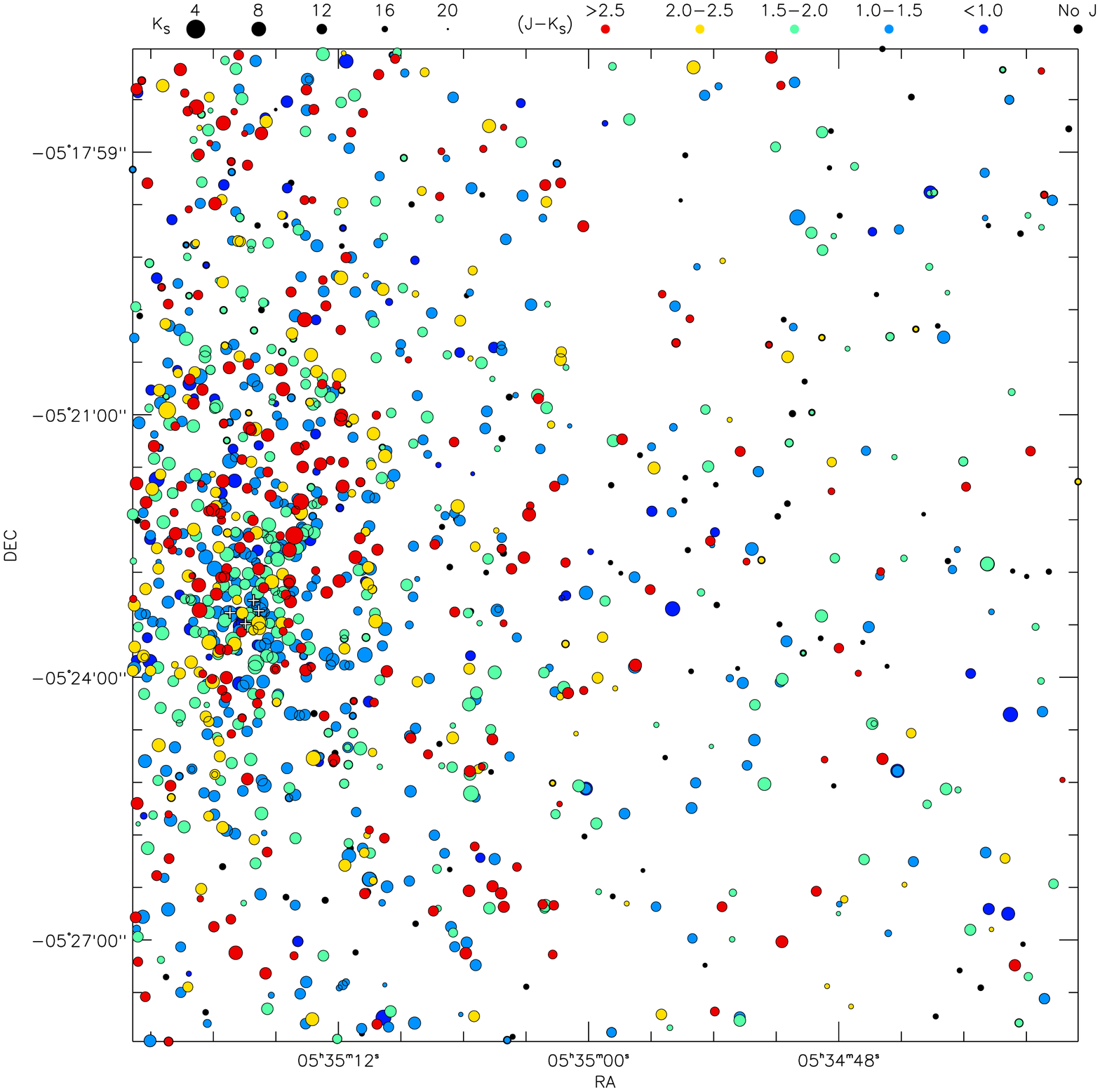}
 \caption{Same as Figure~\ref{fig:findingmap}, for Field 5.}
\end{figure}
\clearpage

%%%
%FIGURE\ 14
%
\begin{figure}
%\epsscale{}
 \plotone{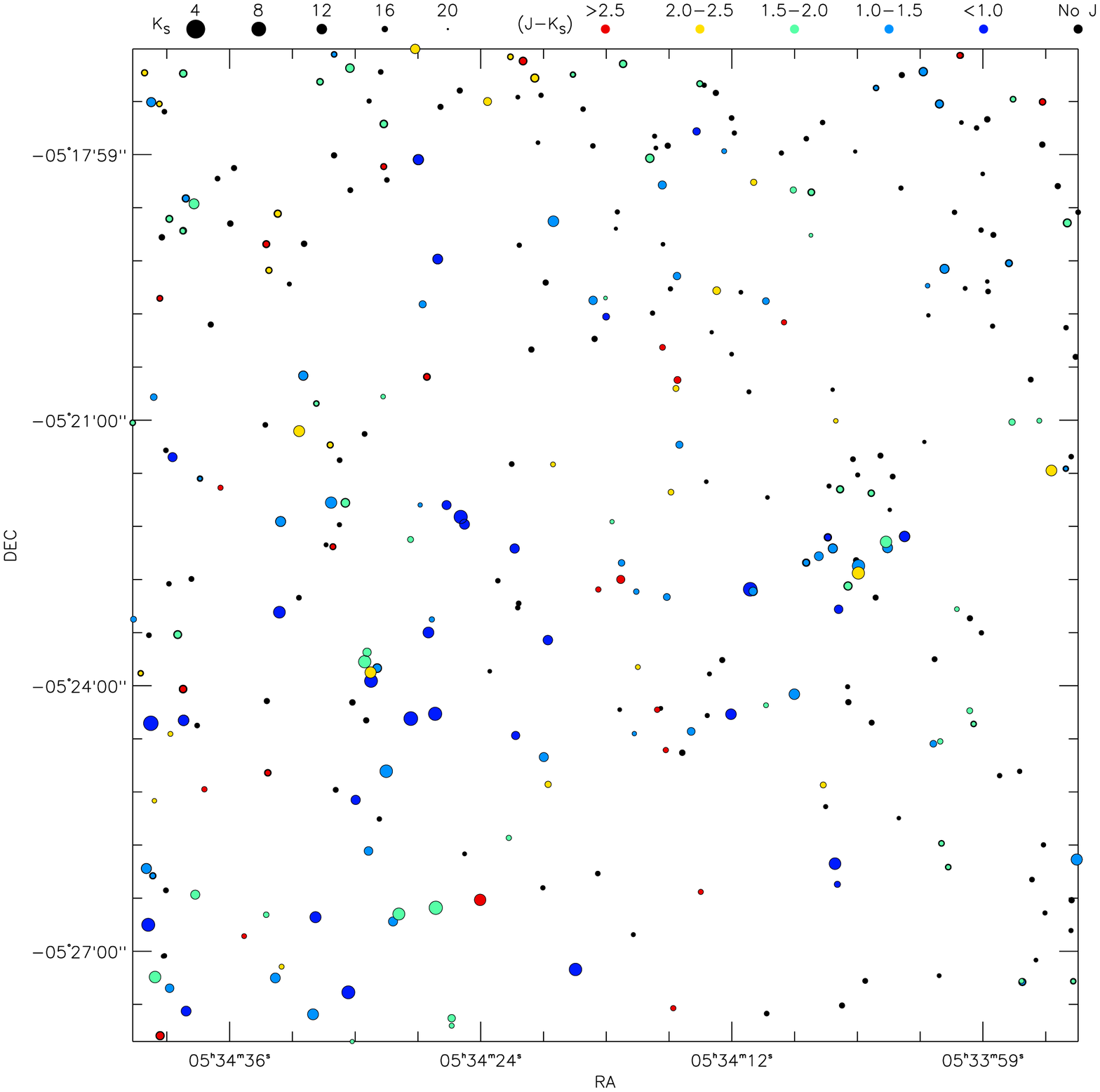}
 \caption{Same as Figure~\ref{fig:findingmap}, for Field 6.}
\end{figure}
\clearpage

%%%
%FIGURE\ 15
%
\begin{figure}
%\epsscale{}
 \plotone{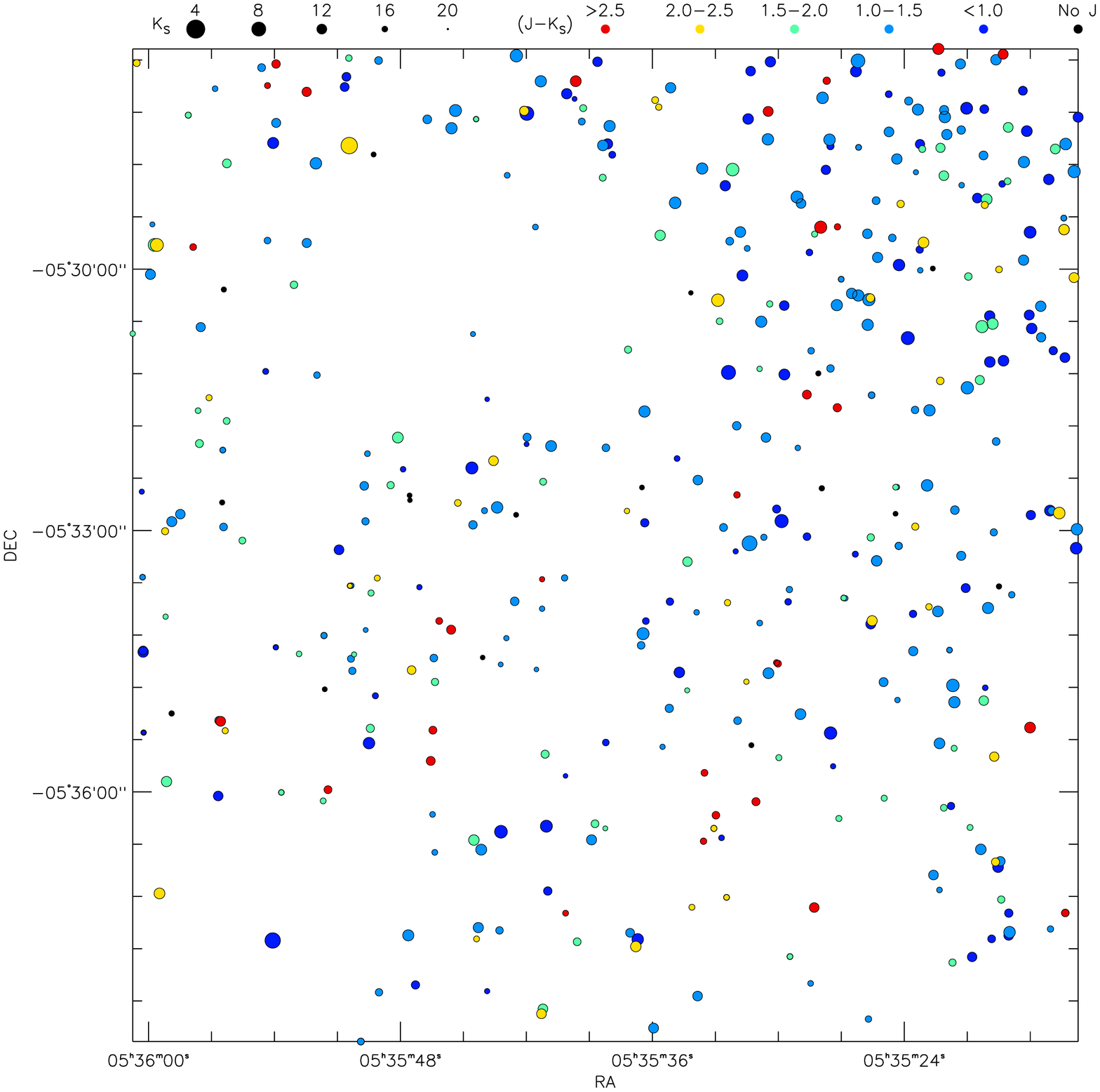}
 \caption{Same as Figure~\ref{fig:findingmap}, for Field 7.}
\end{figure}
\clearpage

%%%
%FIGURE\ 16
%
\begin{figure}
%\epsscale{}
 \plotone{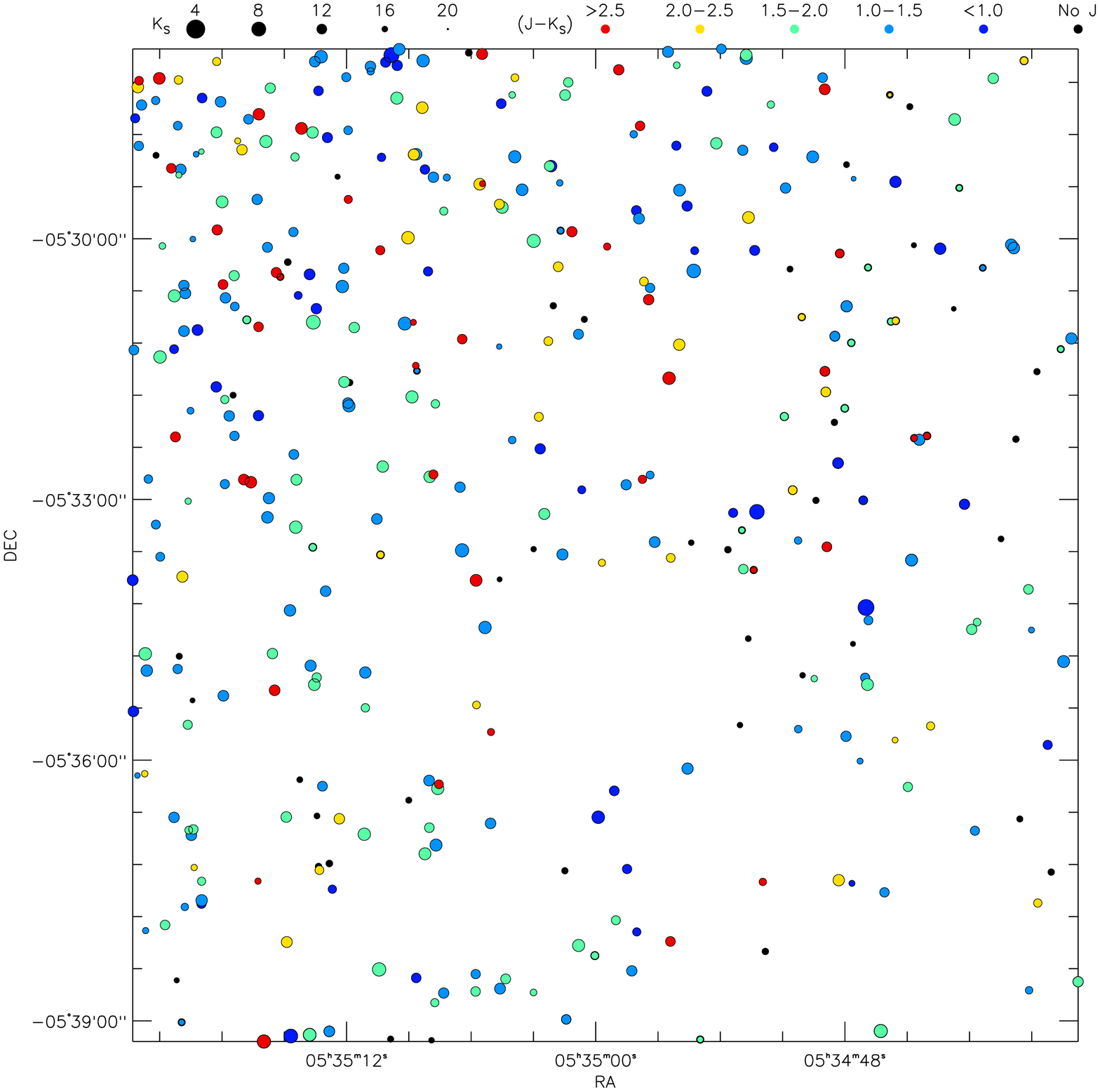}
 \caption{Same as Figure~\ref{fig:findingmap}, for Field 8.}
\end{figure}
\clearpage

%%%
%FIGURE\ 17
%
\begin{figure}
%\epsscale{}
 \plotone{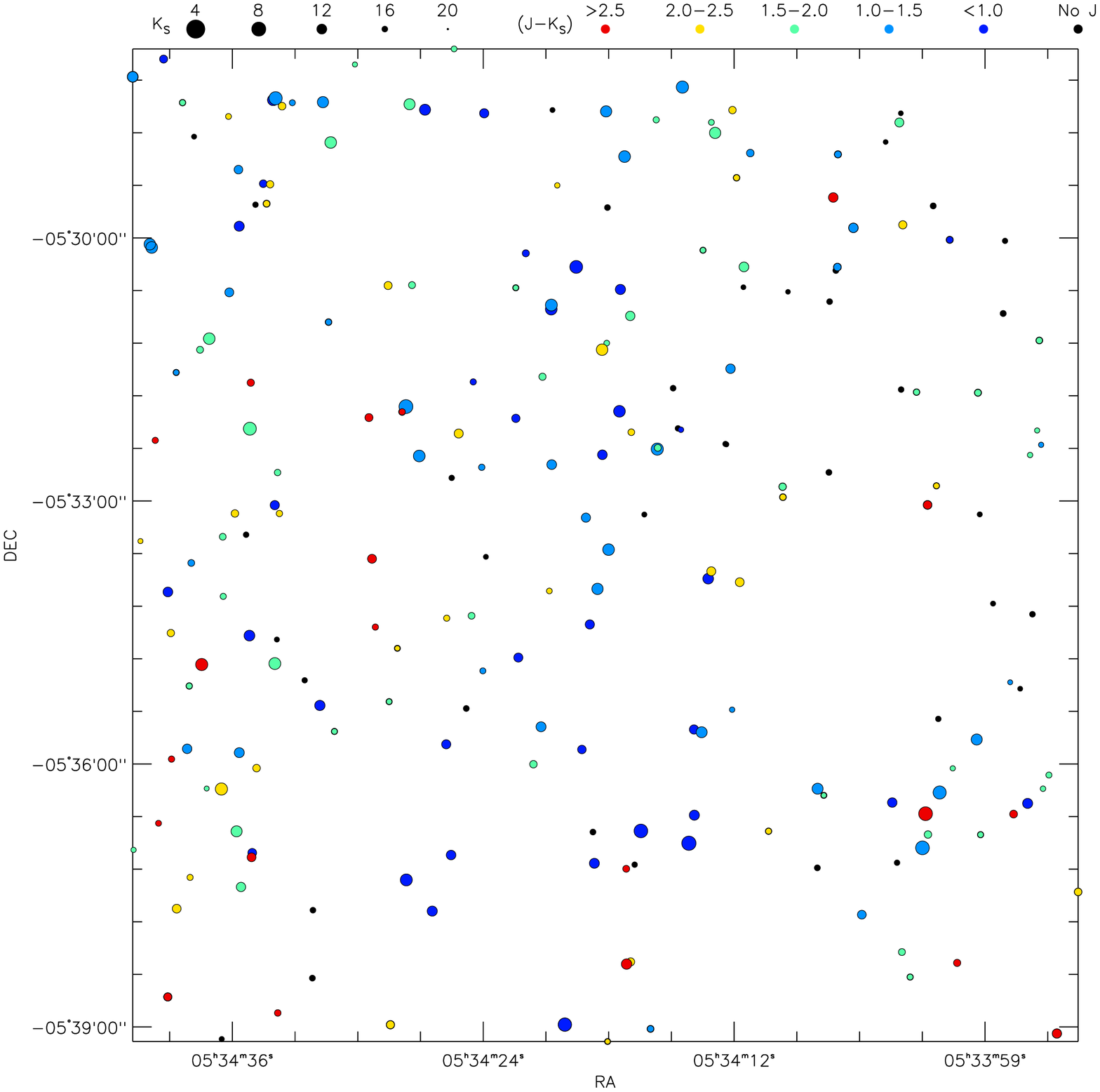}
 \caption{Same as Figure~\ref{fig:findingmap}, for Field 9.}
\end{figure}
\clearpage

%%%
%FIGURE\ 18
%
\begin{figure}
%\epsscale{}
 \plotone{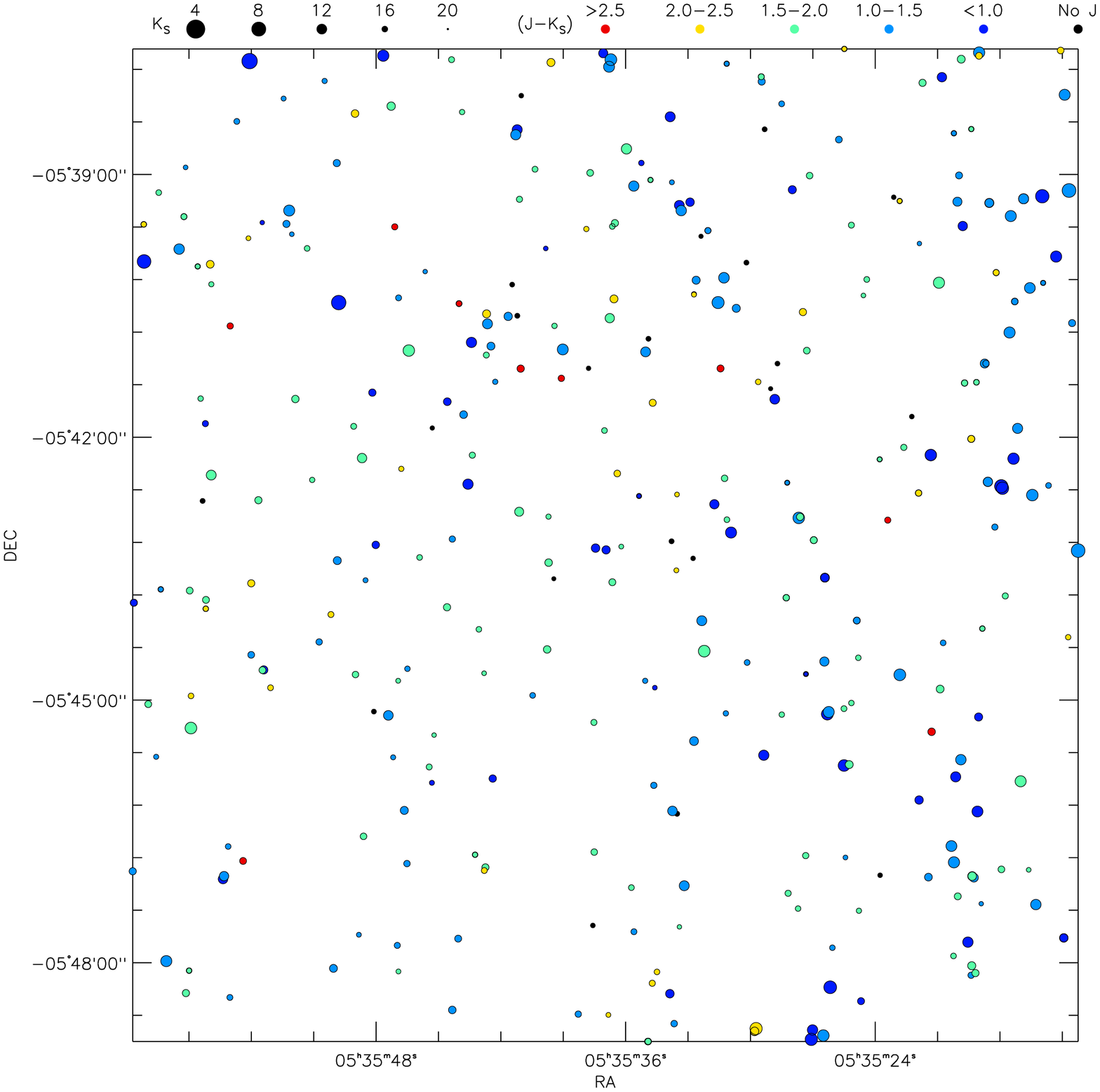}
 \caption{Same as Figure~\ref{fig:findingmap}, for Field 10.}
\end{figure}
\clearpage

%%%
%FIGURE\ 19
%
\begin{figure}
%\epsscale{}
 \plotone{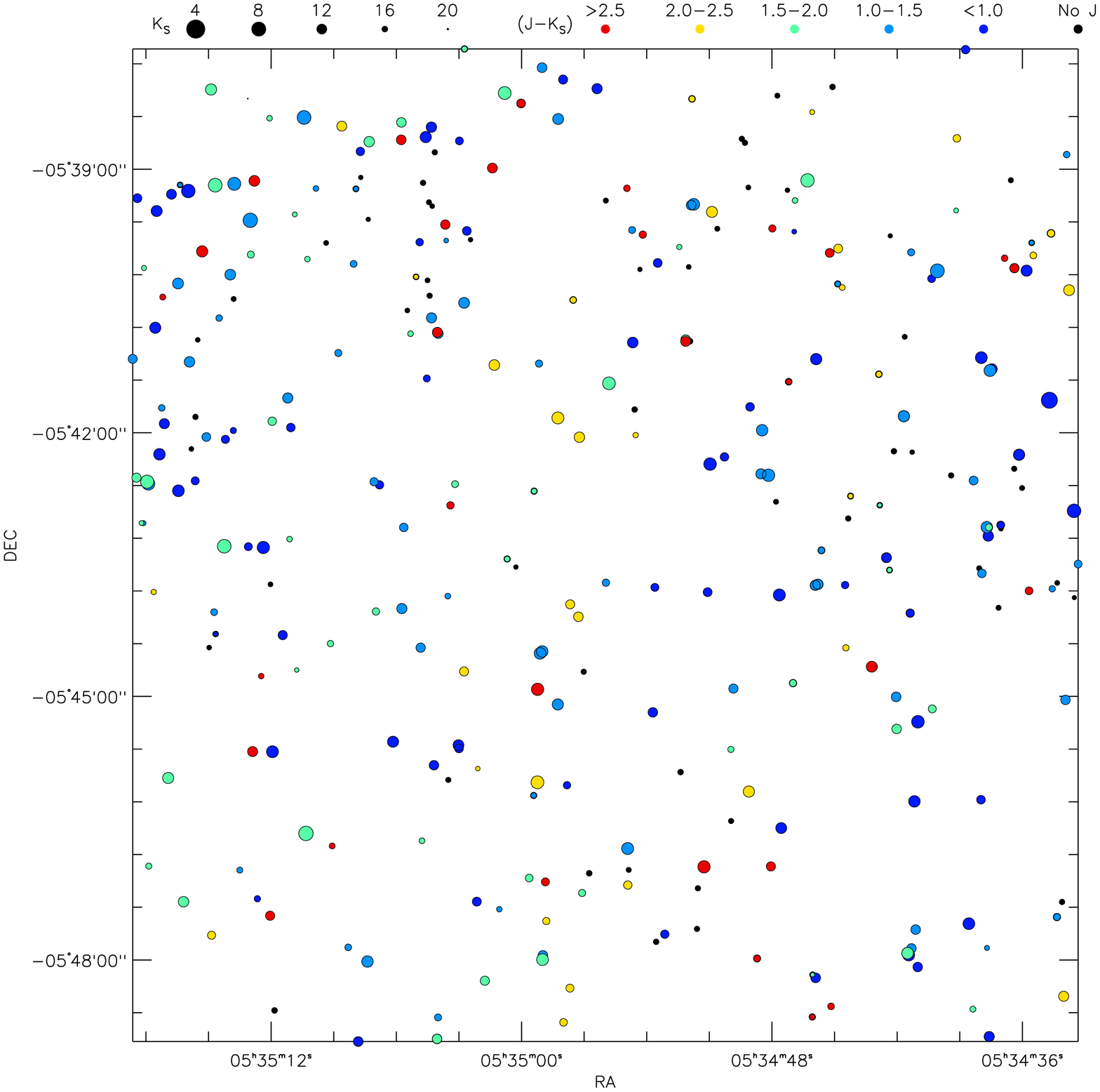}
 \caption{Same as Figure~\ref{fig:findingmap}, for Field 11.}
\end{figure}
\clearpage

%%%
%FIGURE\ 20
%
%\placefigure{fig:ccd}
\begin{figure*}
\centering
\includegraphics[width=.4\linewidth]{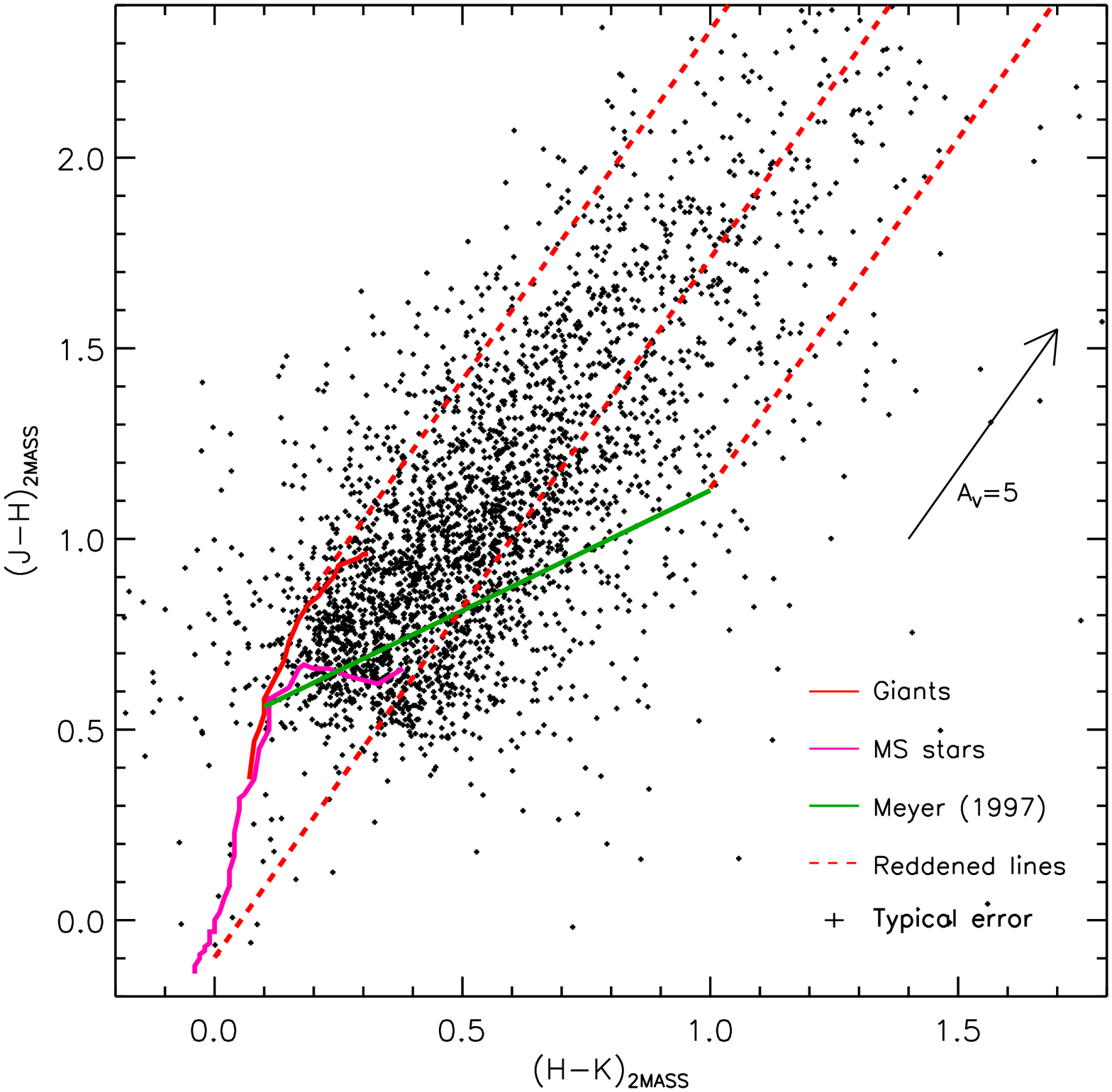}
\includegraphics[width=.4\linewidth]{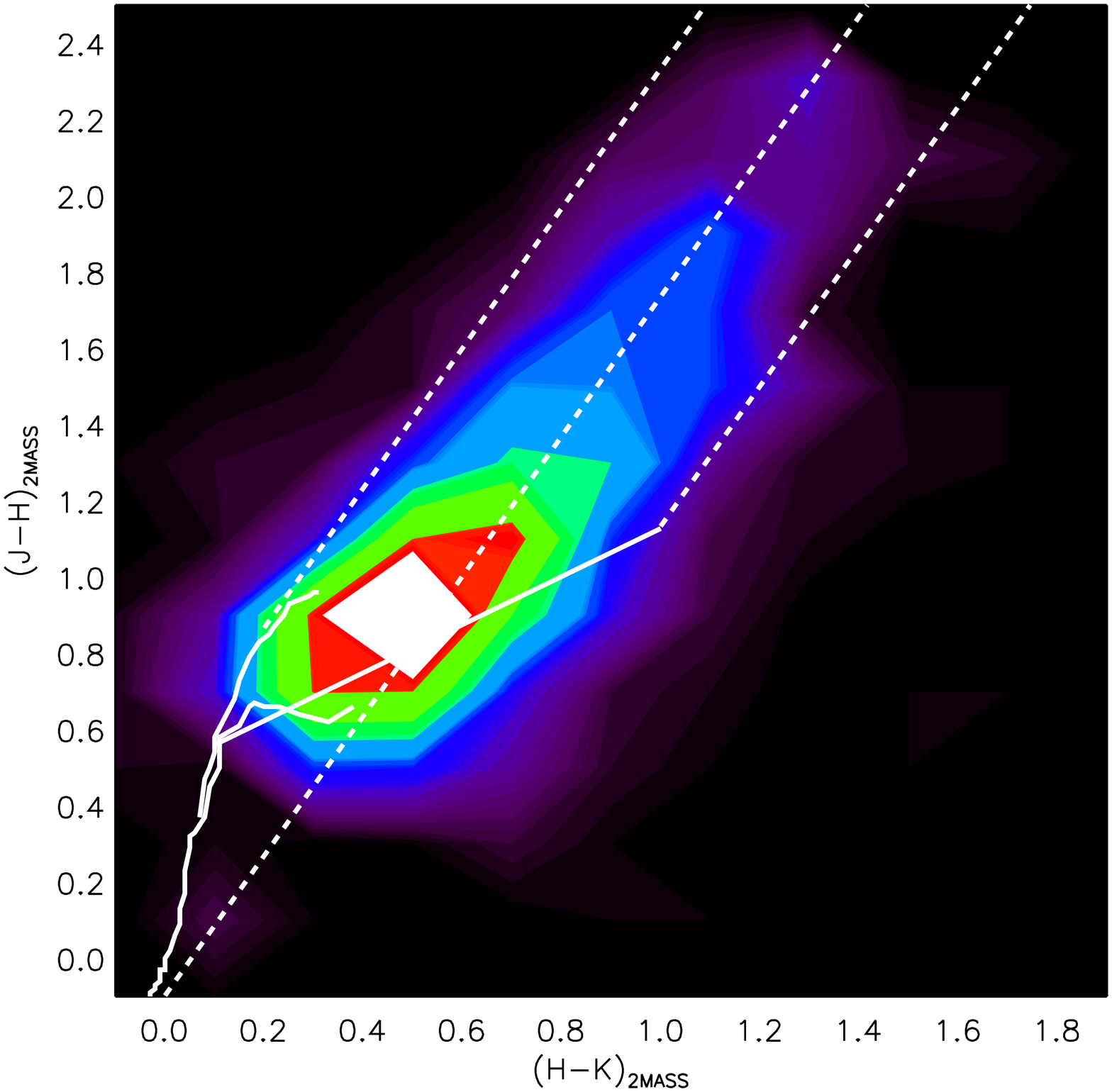}
\includegraphics[width=.4\linewidth]{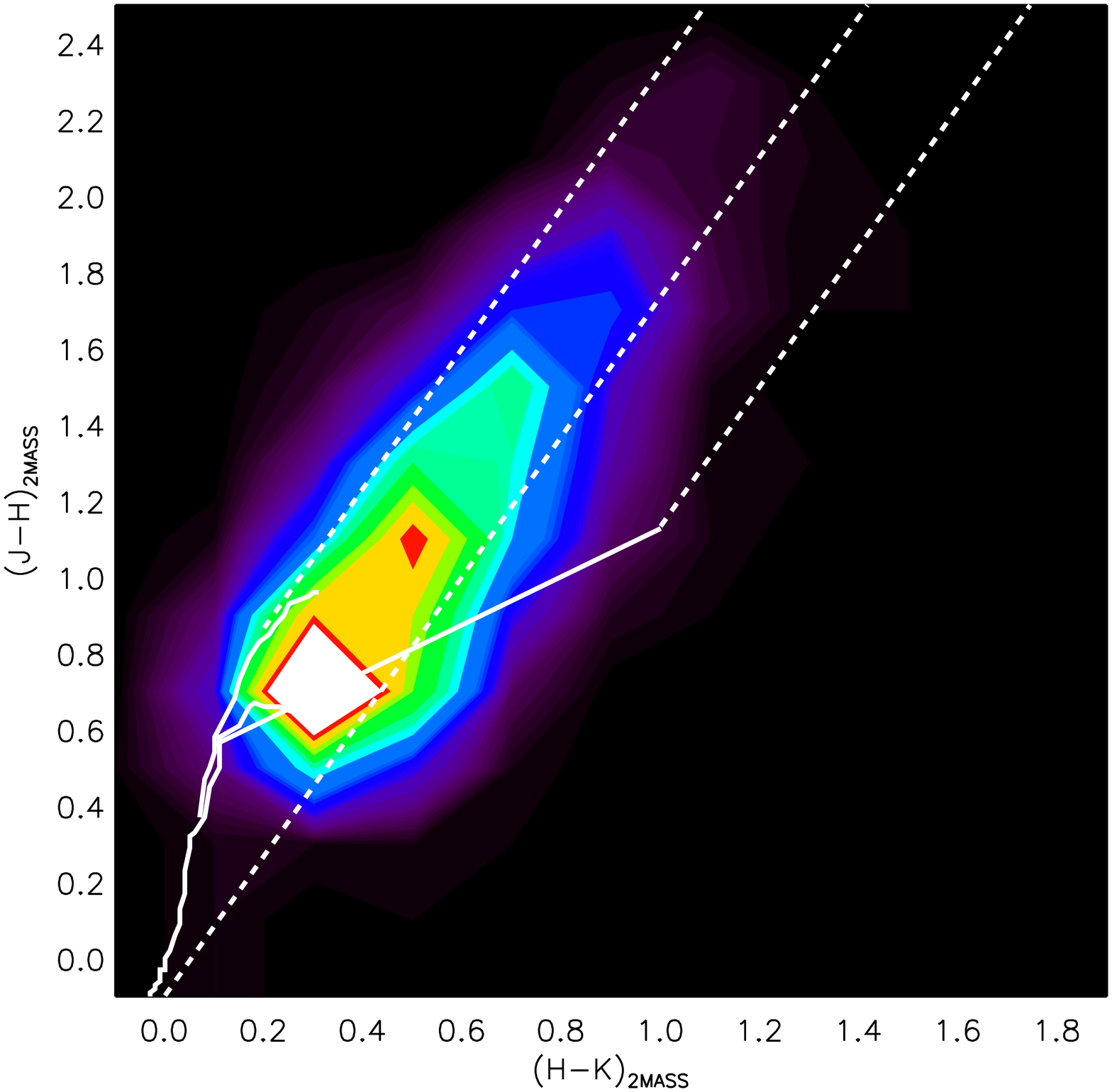}
\includegraphics[width=.4\linewidth]{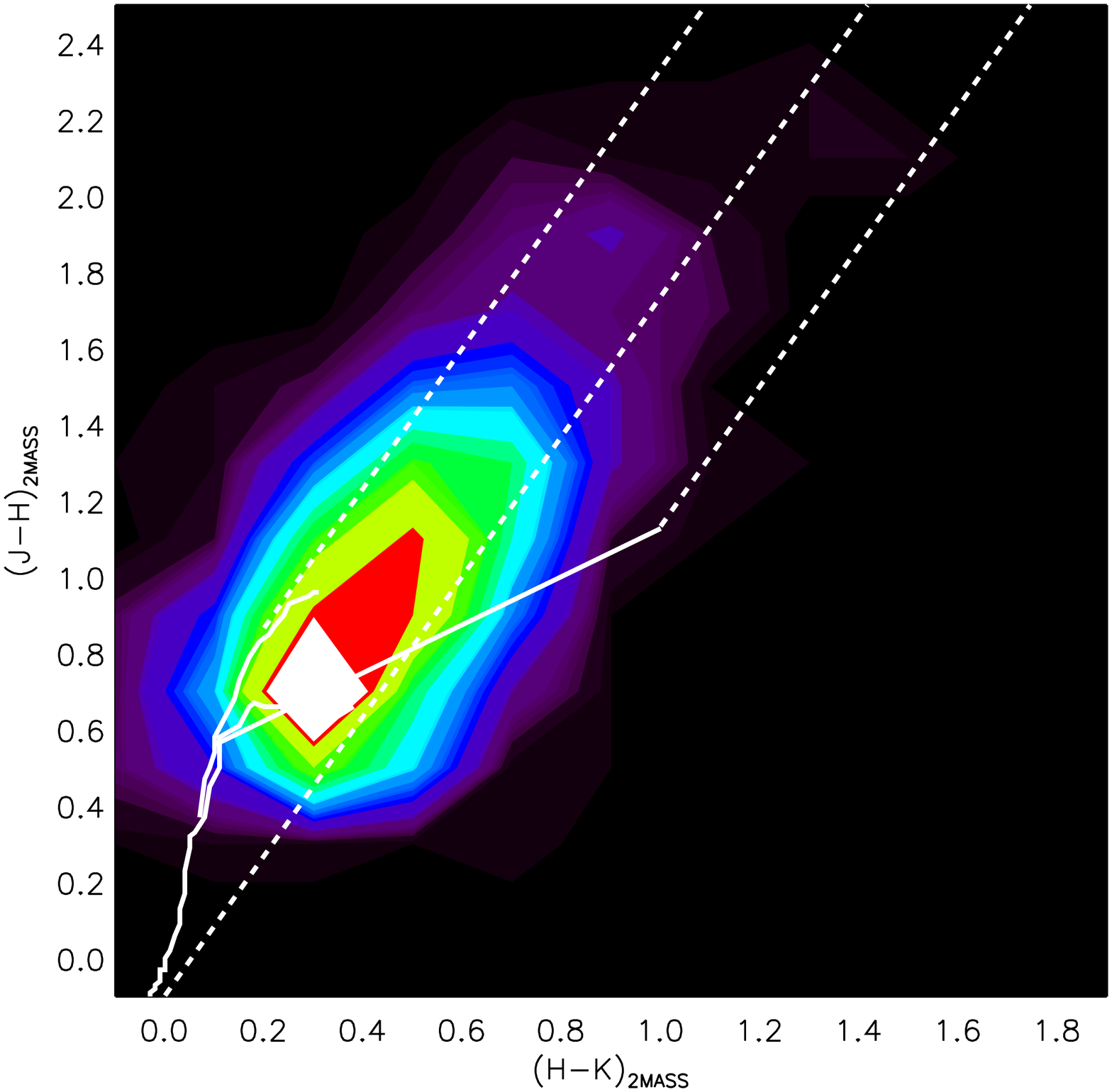}\\
\caption{\textit{a) top-left panel}: Near-infrared color-color diagrams for selected point-like sources in our catalog detected in all 3 bands (see text for the selection criteria). Also shown are the main-sequence (magenta solid line), the giant-branch (red solid line) sequence
%, the isochrone at 1Myr for stellar masses given by \citet{Sie00} (cyan solid line), the isochrone for sub--stellar masses given by \citet{CBAH00}  (cyan dashed line) 
 and the CTTs locus by \citet{Meyer97} (green solid line). The dotted lines are parallel to the standard reddening vector. \textit{b)\ (top-right panel)} - Hess diagrams of the color-color diagram relative to the inner region.   \textit{c)\ bottom-left panel}:  Similar to panel \textit{b)}, for the median region. \textit{d)\ bottom-right panel}:  Similar to panel \textit{b)} for the outer region.}\label{fig:ccd}
\end{figure*}
\clearpage

%%%
%FIGURE\ 21
%
%\placefigure{fig:hr}
\begin{figure*}
 \epsscale{1}
 \plottwo{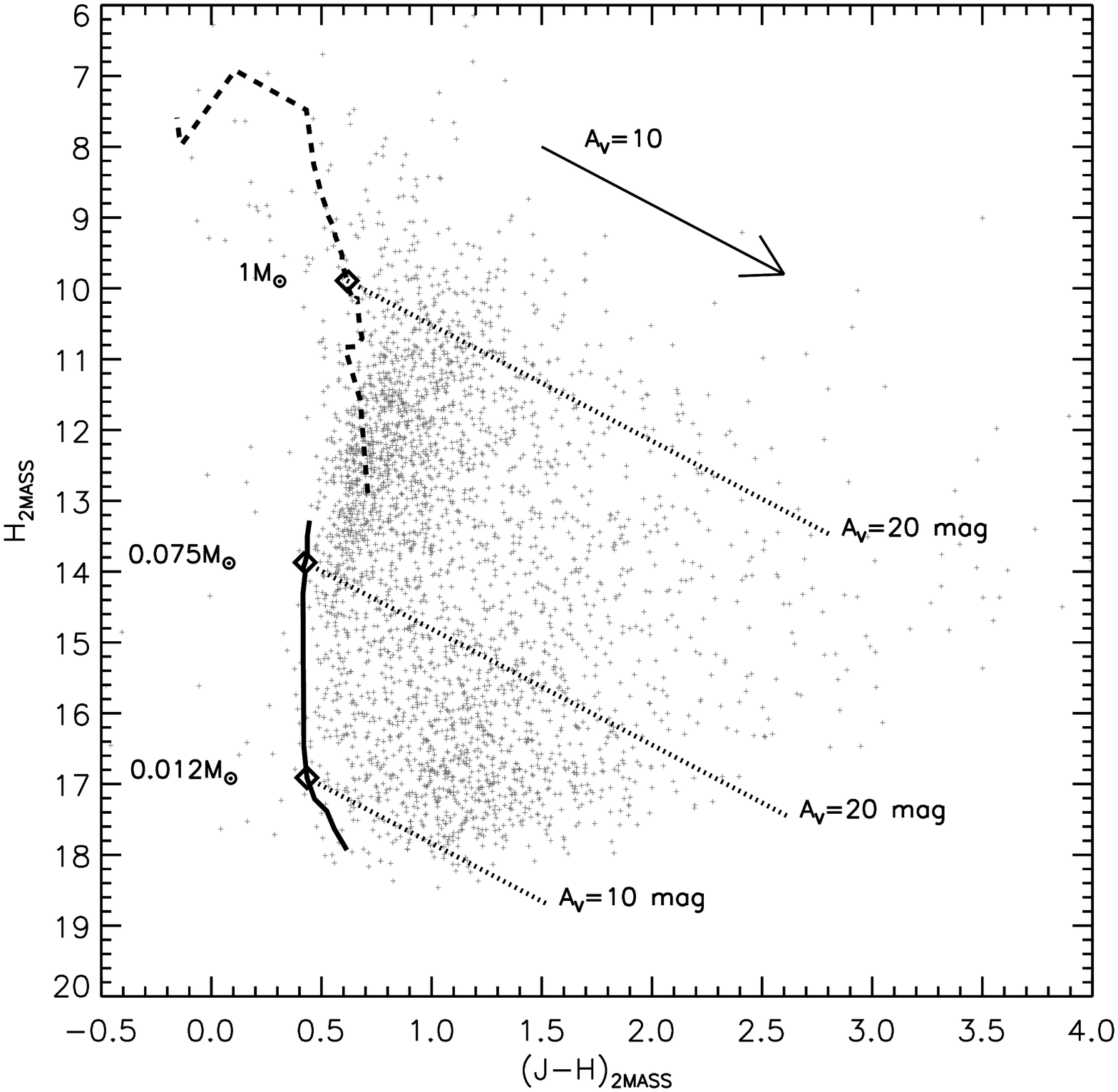}{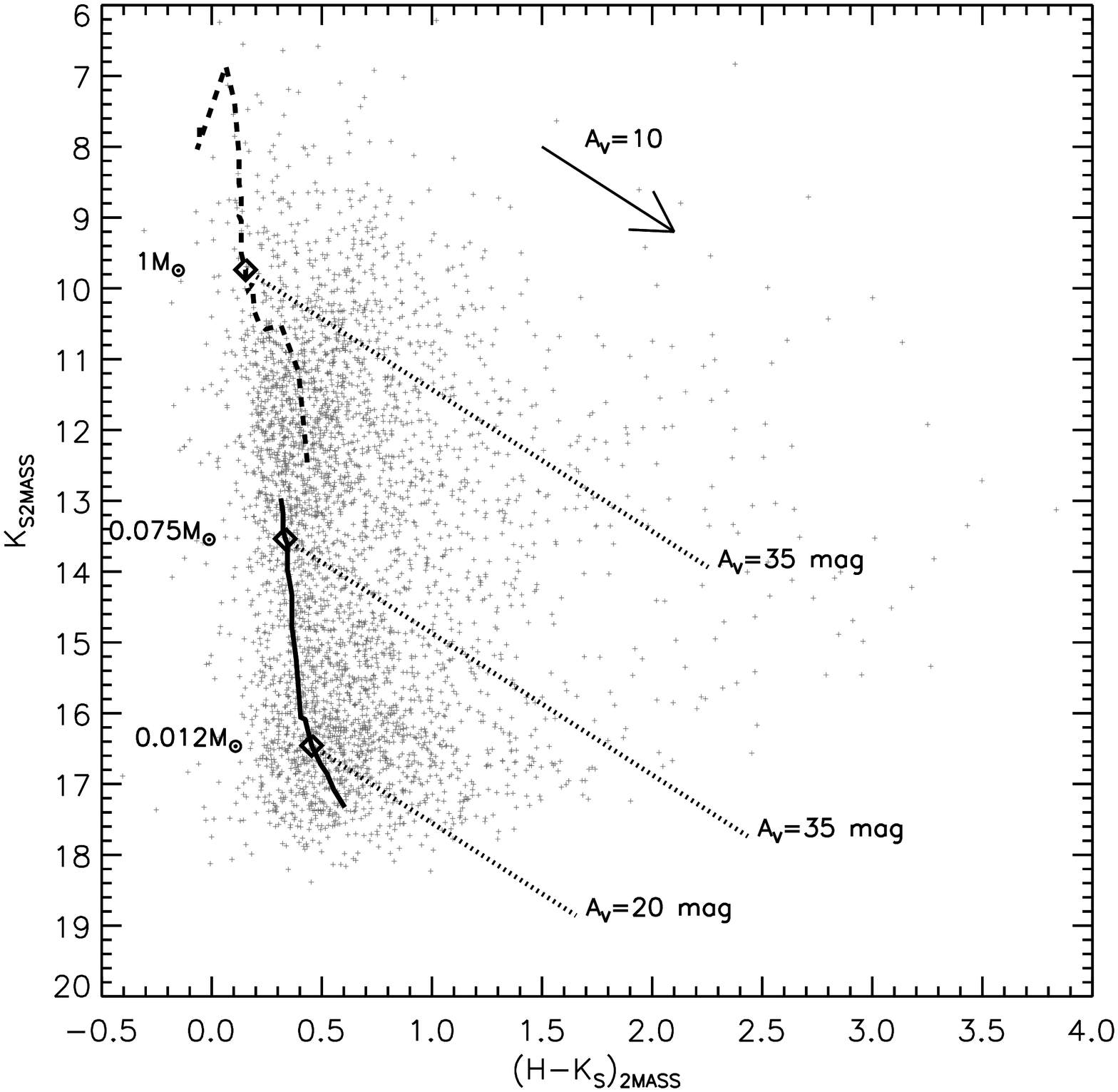}
 \caption{Infrared color-magnitude diagrams for  selected point sources in our catalog detected in all 3 bands, with 1Myr isochrones, assuming a distance $d=414$~pc ($DM=8.01$). To derive these isochrones, we used \citep{CBAH00} combined to the \citep{Sie00} model, as discussed in Section\ \ref{sec:cmd}. The $A_V$=10~mag reddening vector is also shown. The diamond symbols represent the positions of a $M=0.012$~M$_\odot$ (deuterium burning limit) object, a $M=0.075$~M$_\odot$ (hydrogen burning limit) object and a $M=1$~M$_\odot$ star from bottom to top respectively.}\label{fig:hr}
\end{figure*}
\clearpage

%%%
%FIGURE\ 22
%
%\placefigure{fig:cmhess}
\begin{figure*}
\centering
\includegraphics[width=.4\linewidth]{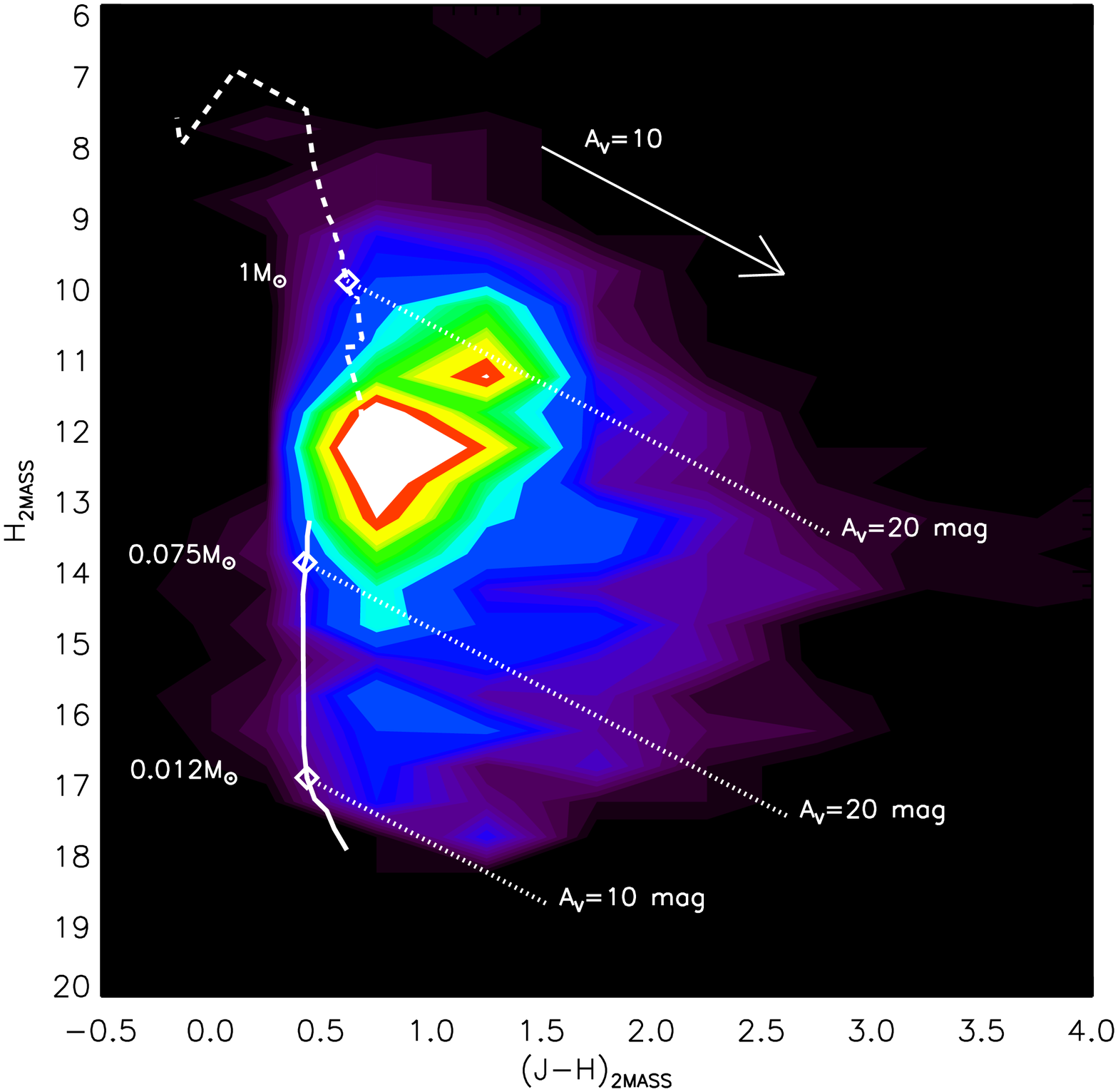}
\includegraphics[width=.4\linewidth]{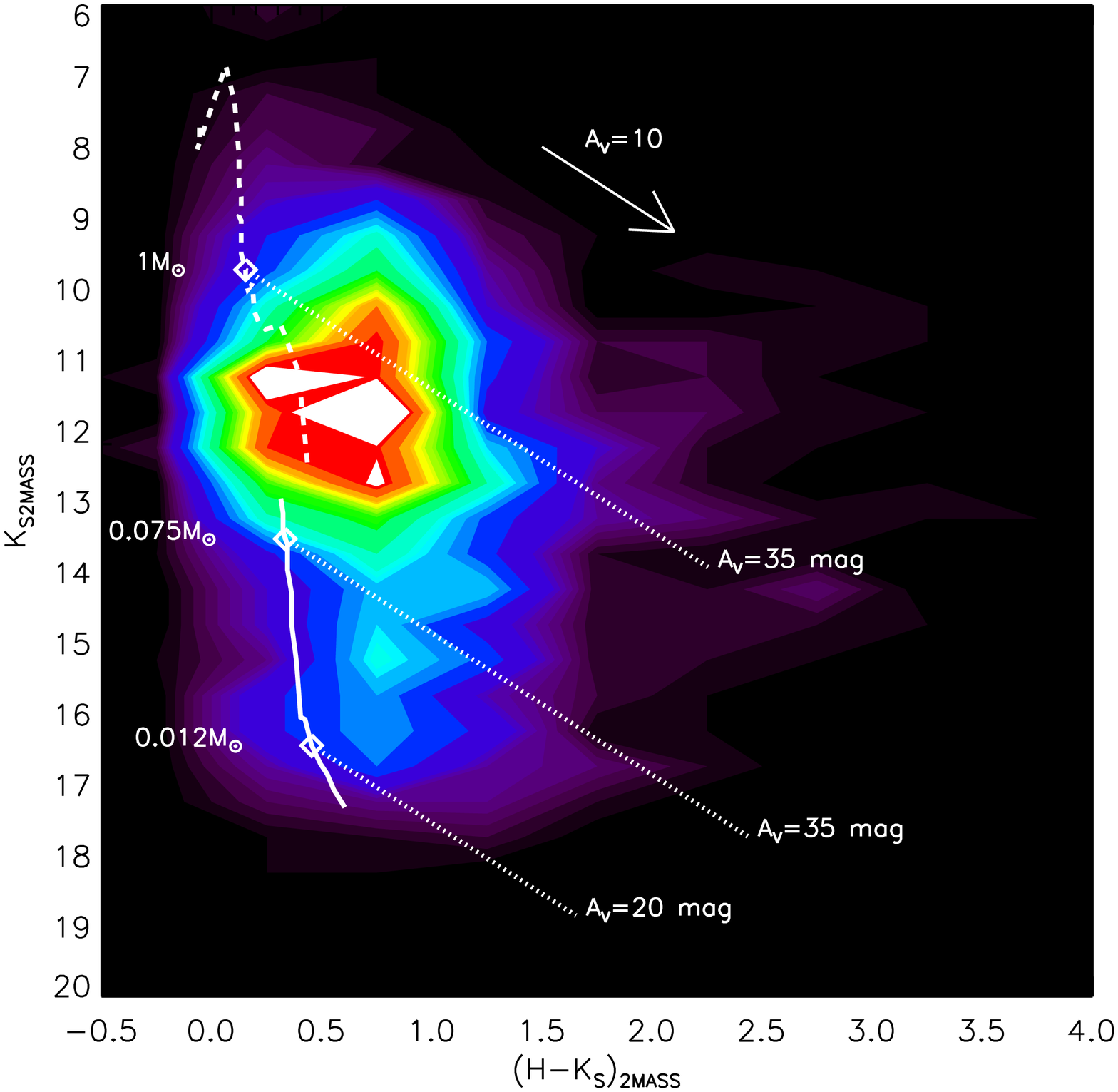}\\
\includegraphics[width=.4\linewidth]{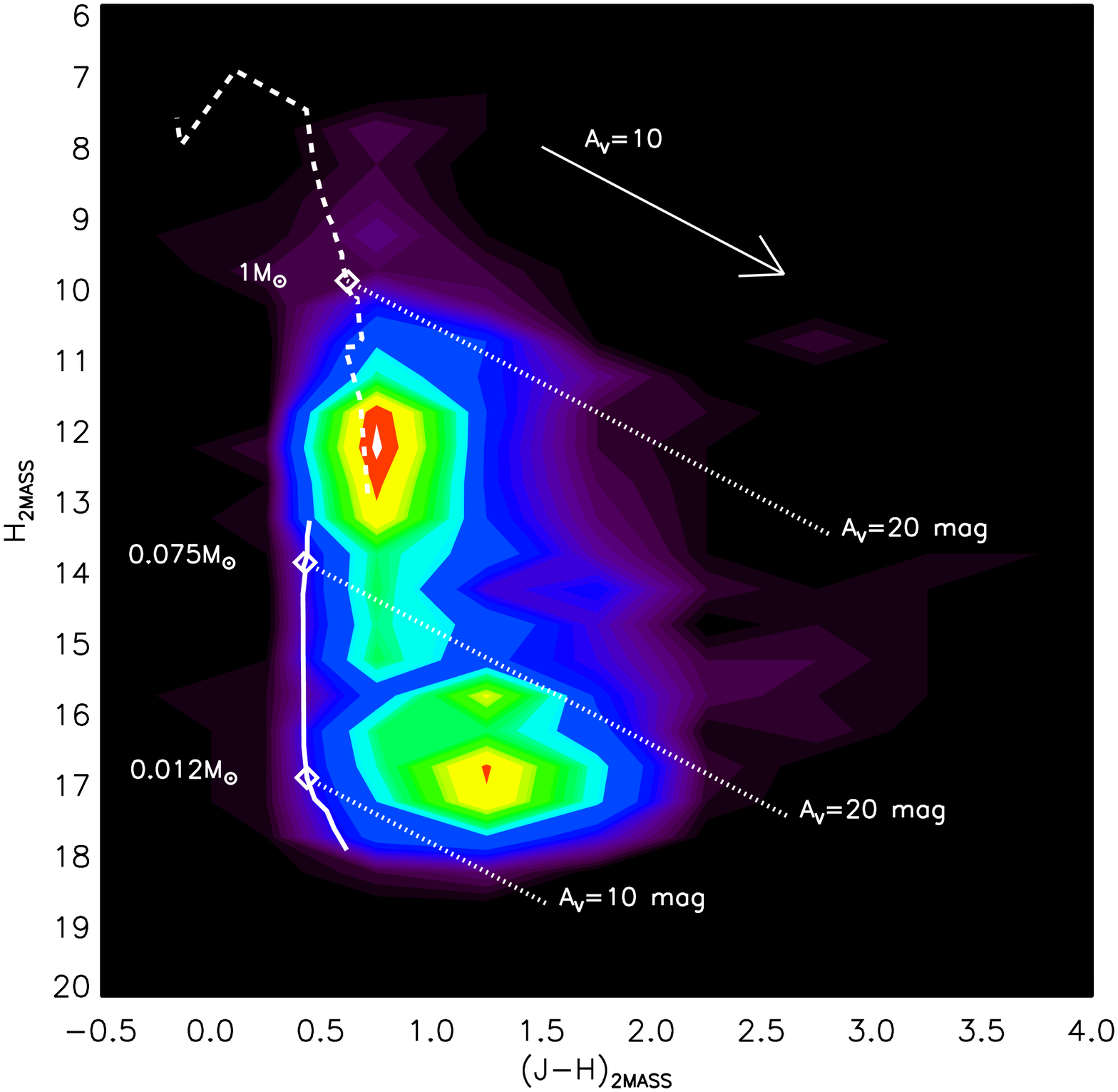}
\includegraphics[width=.4\linewidth]{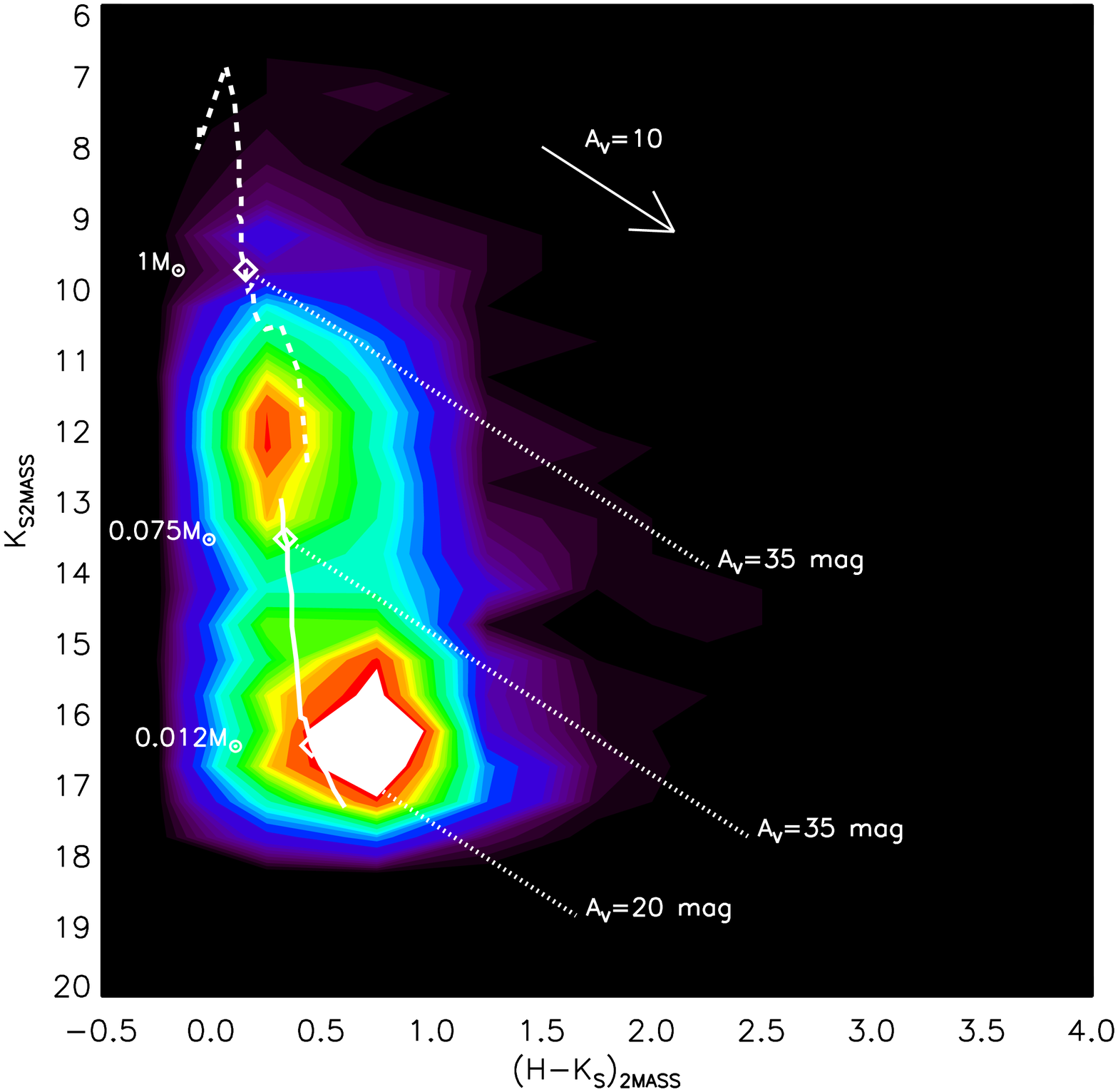}\\
\includegraphics[width=.4\linewidth]{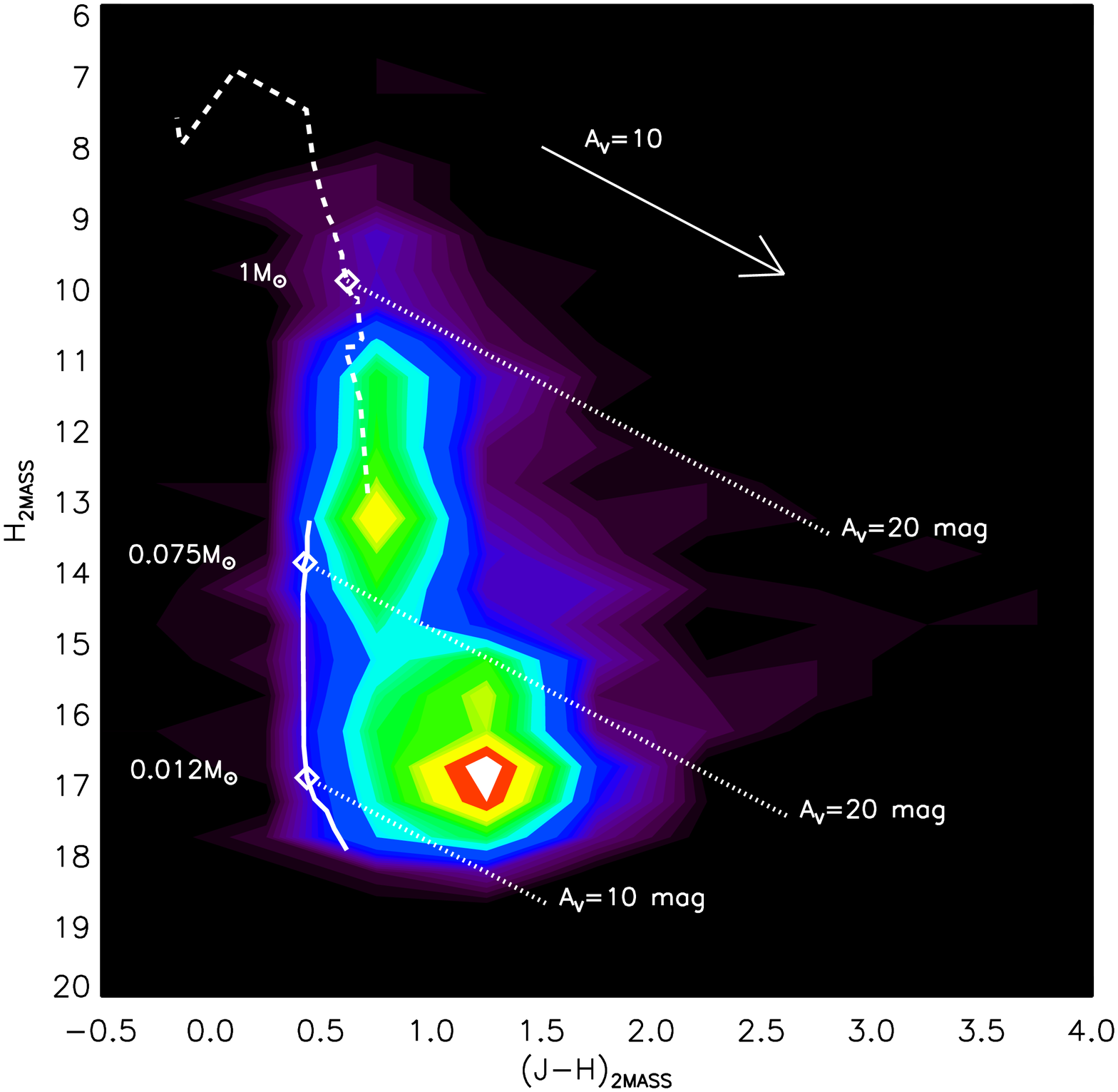}
\includegraphics[width=.4\linewidth]{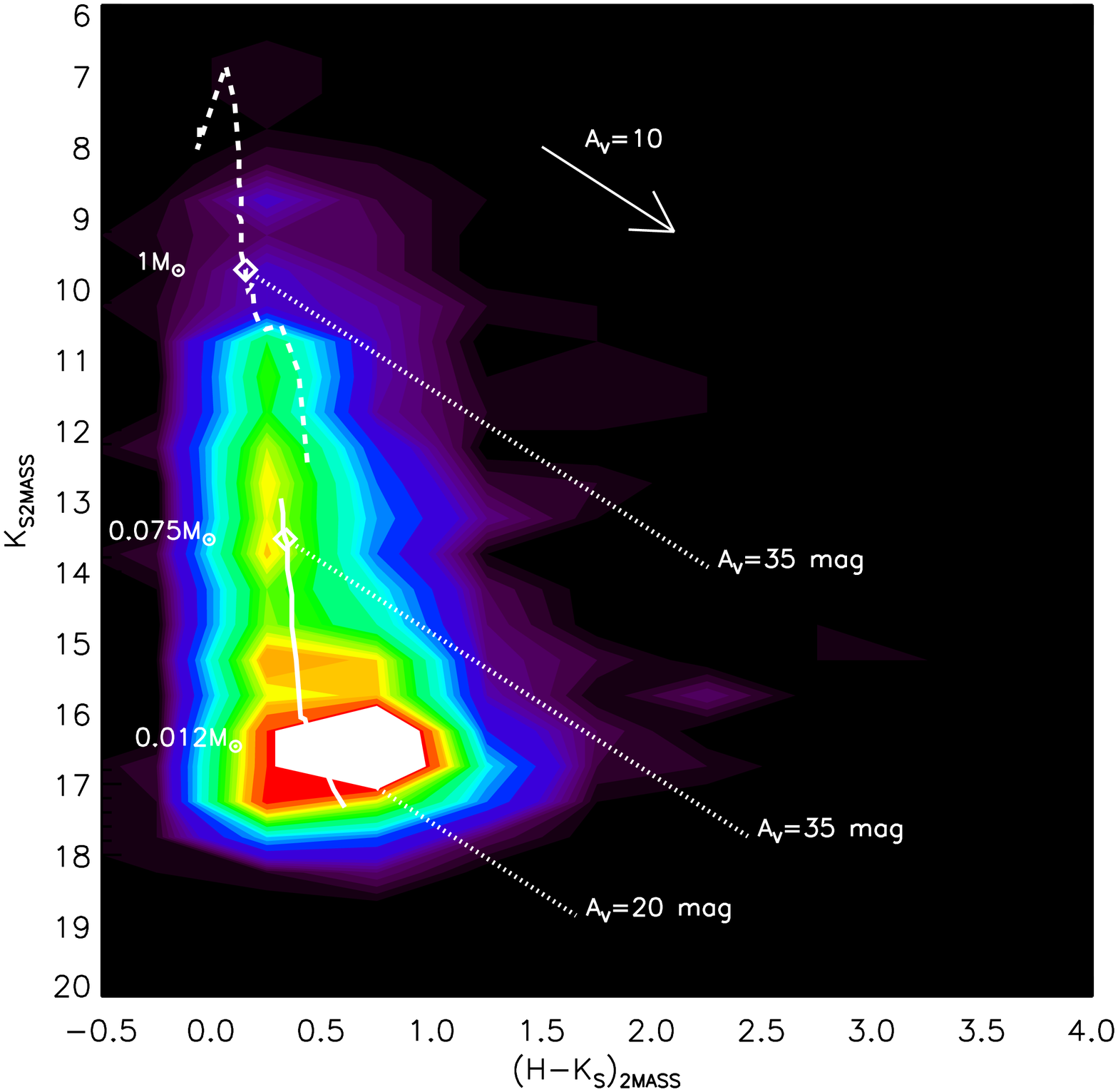}\\
\caption{Hess format for the \jb,\jb-\hb ~(left panels) and \kb,\hb-\kb ~(right panels) color-magnitude diagrams. Line styles and symbols are the same as in Figure\ \ref{fig:hr}. From top to bottom, the diagrams for the inner ($r<6.7^\prime$), medium ($6.7^\prime<r<14.3^\prime$) and outer ($14.3^\prime<r<27.2^\prime$) regions are shown. }\label{fig:cmhess}
\end{figure*}
\clearpage

%%%
%FIGURE\ 23
%
%\placefigure{fig:lfcomplfig}
\begin{figure*}
\centering
\includegraphics[width=.32\linewidth]{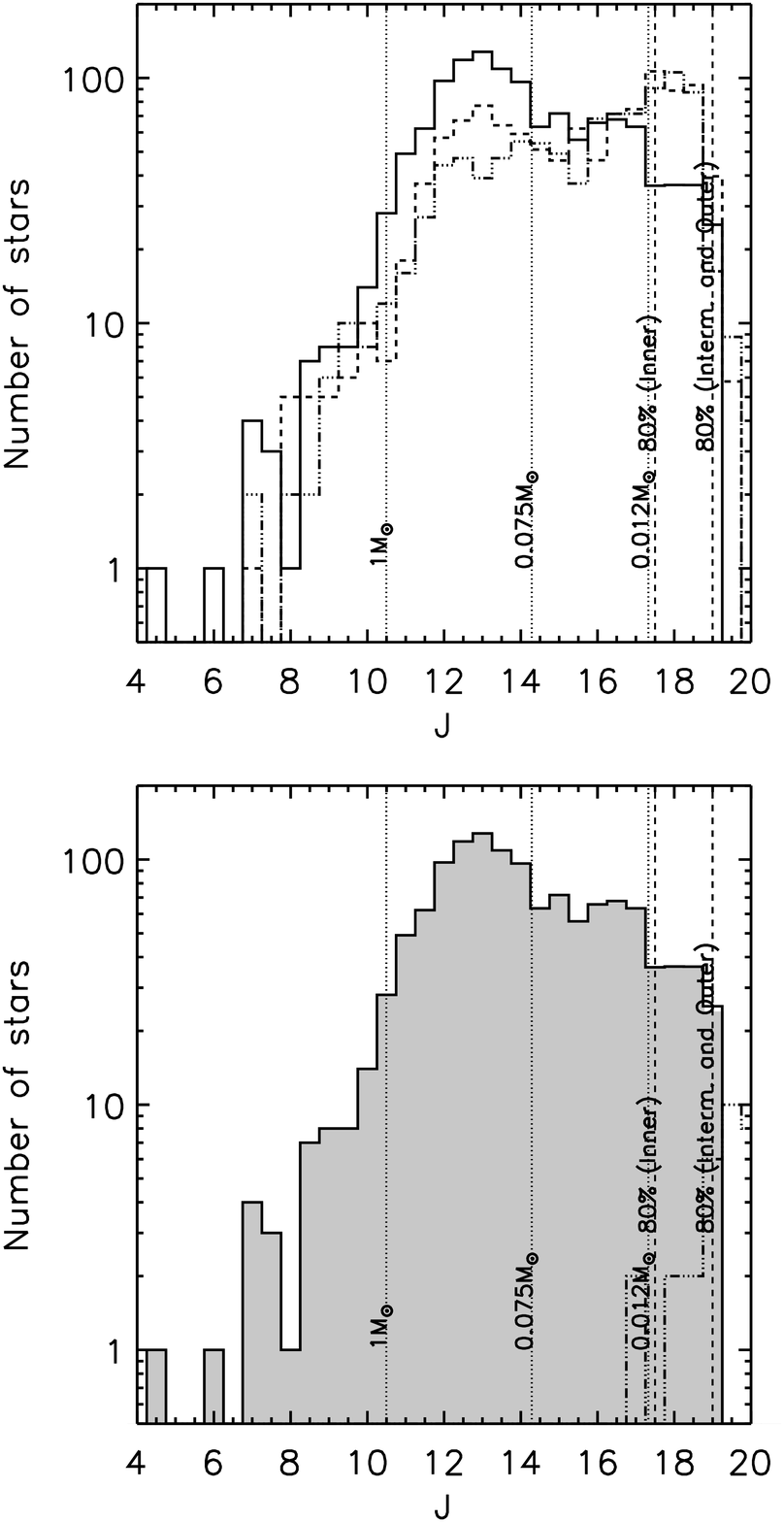}
\includegraphics[width=.32\linewidth]{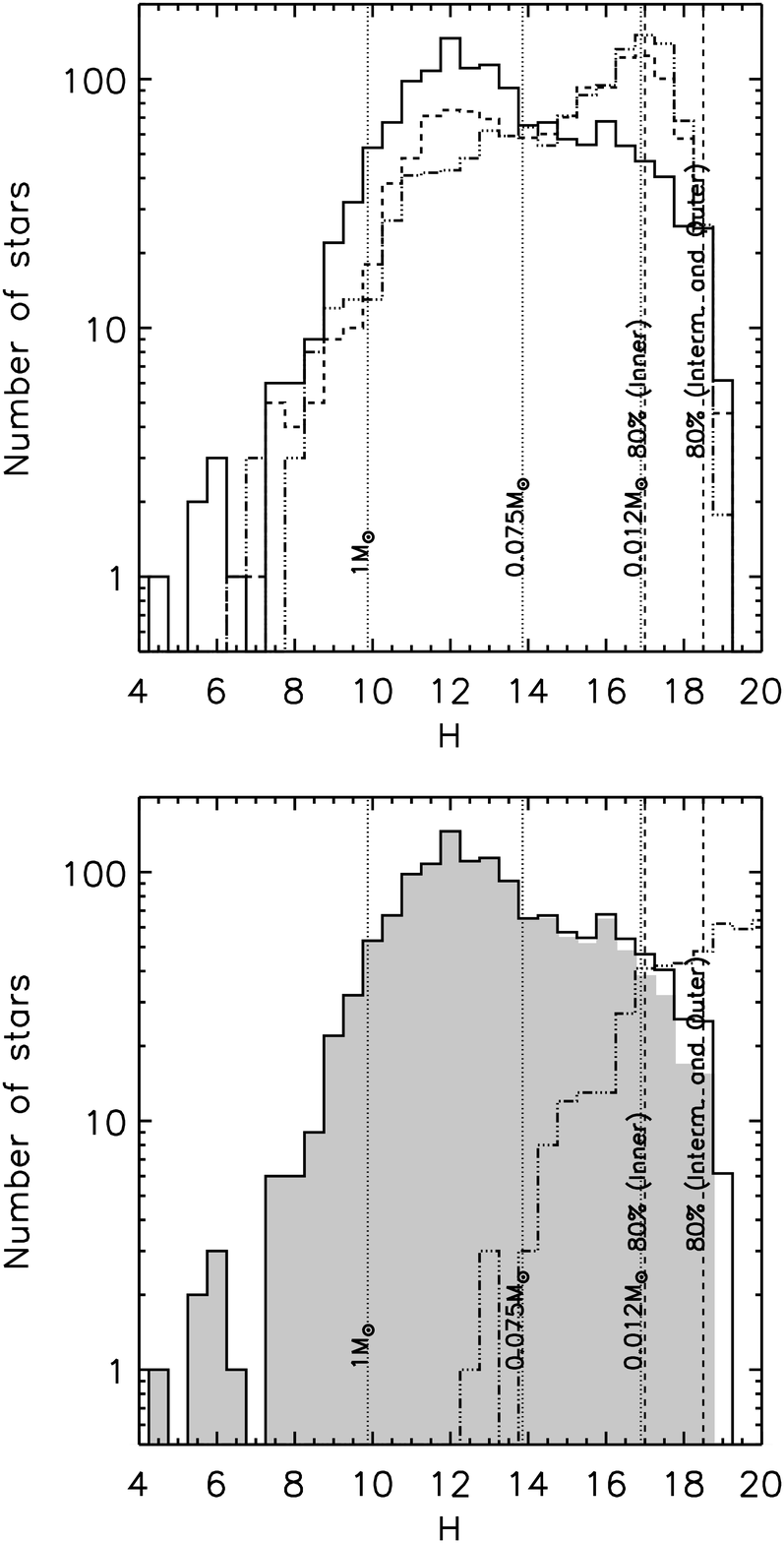}
\includegraphics[width=.32\linewidth]{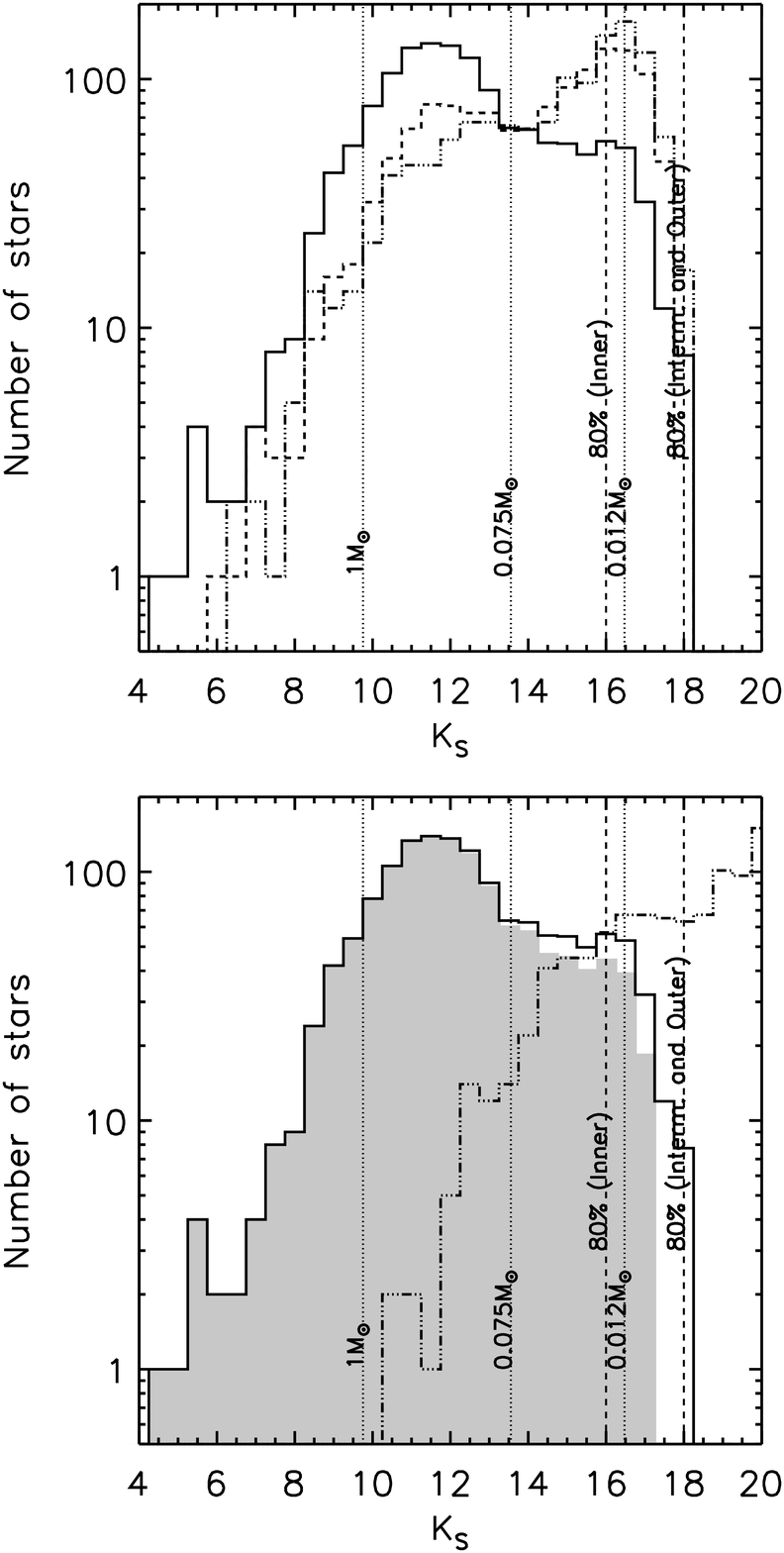}
\caption{\textit{Top panels}: \jb-band (\textit{left}), \hb-band (\textit{center}) and \kb-band (\textit{right}) luminosity functions for the ONC, corrected for completeness. The solid line is relative to the inner region,  the dashed line to the intermediate region and the dotted-dashed lines to the outer region. \textit{Bottom panels}: for each bandpass, luminosity function of the inner region (solid line), of the outer region shifted by $A_V=33$~mag of extinction (dot-dashed line), and difference of the two (gray area). In each plot, we indicate with vertical lines the magnitudes corresponding to stars 1~Myr old at 414~pc with 1~$M_\odot$, 0.075~$M_\odot$ and 0.012~$M_\odot$ respectively, without interstellar reddening. The vertical dashed line stands for the 80\% sensitivity level in the outer region.}\label{fig:lfcomplfig}
\end{figure*}
\clearpage

%%%
%TABLE\ 1
%
\begin{deluxetable}{lccc}
\tabletypesize{\scriptsize}
%\rotate
\tablecaption{ISPI Broad Band Filters \label{Tab_ISPI_Filters}}
\tablewidth{0pt}
\tablehead{
\colhead{Filter} & \colhead{Central wavelength} & \colhead{$\Delta\lambda (80\%$)}\\
                 &             ($\mu$m)           &    ($\mu$m) }
\startdata
       \jb &     1.250 &  1.176 - 1.322 \\
       \hb &     1.635 &  1.500 - 1.770 \\
       \kb &    2.150 &  1.992 - 2.300 \\
\enddata\
%\tablecomments{}
\end{deluxetable}
\clearpage

%%%
%TABLE\ 2
%
\begin{deluxetable}{cccccc}
\tabletypesize{\scriptsize}
%\rotate
\tablecaption{Log of the ISPI observations\label{Tab_ISPI-logbook}}
\tablewidth{0pt}
\tablehead{
\colhead{Field} & \colhead{Night} & \colhead{RA (2000.0)} & \colhead{DEC (2000.0)} & \colhead{Airmass} & \colhead{Comment}
}
\startdata
1 & A & 05:35:38.37 & -5:12:50.7 & 1.144 - 1.291 &   \\
2 & A & 05:34:59.63 & -5:12:57.6 & 1.123 - 1.139 &    \\
3 & A & 05:34:18.52 & -5:12:59.6 & 1.123 - 1.139 & no \jb-band 30s exposure   \\
4 & A & 05:35:37.05 & -5:22:38.6 & 1.309 - 1.158 &    \\
  & B & 05:35:36.96 & -5:22:50.6 & 1.309 - 1.158 & field 12 in the database   \\
5 & A & 05:34:59.08 & -5:22:28.3 & 1.319 - 1.938 &    \\
  & B & 05:34:59.21 & -5:22:55.0 & 1.225 - 1.498 & field 13 in the database   \\
6 & A & 05:34:17.98 & -5:22:28.3 & 1.319 - 1.938 &    \\
  & B & 05:34:18.01 & -5:22:55.7 & 1.225 - 1.498 & field 14 in the database   \\
7 & B & 05:35:38.16 & -5:33:03.8 & 1.218 - 1.410 &    \\
8 & B & 05:34:59.82 & -5:33:18.6 & 1.104 - 1.211 &    \\
9 & B & 05:34:18.50 & -5:33:21.4 & 1.104 - 1.211 &    \\
10 & B & 05:35:37.48 & -5:43:06.9 & 1.110 - 1.215 &   \\
11 & B & 05:34:56.28 & -5:43:09.2 & 1.110 - 1.215 &   \\
\enddata
\end{deluxetable}
\clearpage

%%%
%TABLE\ 3
%
\begin{deluxetable}{ccrrrrr}
 \tabletypesize{\scriptsize}
 %\rotate
 \tablecaption{Best estimate for calibration coefficients.\label{tab:fitpar}}
 \tablewidth{0pt}
\tablehead{
 \colhead{Field}  & \colhead{$Z_{H}$} & \colhead{$\sigma_{Z_{H}}$} & \colhead{$\epsilon_{JK}$} & \colhead{$\sigma_{\epsilon_{JK}}$}
 }
 \startdata
        1 &     1.09 &     0.01 &     -0.010 &     0.008\\
        2 &     1.07 &     0.01 &      0.001 &     0.007\\
        3 &     1.08 &     0.04 &      0.00  &     0.02 \\
        4 &     1.13 &     0.01 &     -0.022 &     0.006\\
        5 &     1.11 &     0.01 &     -0.023 &     0.007\\
        6 &     1.03 &     0.02 &     -0.05  &     0.02 \\
\hline
        7 &     0.93 &     0.02 &      0.00  &     0.01 \\
        8 &     0.93 &     0.01 &     -0.023 &     0.008\\
        9 &     0.95 &     0.01 &     -0.02  &     0.01 \\
       10 &     1.09 &     0.02 &     -0.06  &     0.01 \\
       11 &     1.07 &     0.02 &     -0.02  &     0.01 \\
\enddata
\tablecomments{Parameters for field 3 are computed using non-normalized \jb-band magnitudes.}
% \tablenotetext{a}{Sample footnote for table~\ref{tbl-1} that was generated with the deluxetable environment}
% \tablenotetext{b}{Another sample footnote for table~\ref{tbl-1}}
\end{deluxetable}
\clearpage

%%%
%TABLE\ 4
%
\begin{deluxetable}{lr}
\tablewidth{0pc}
\tablecaption{Source Summary\label{tab:catalog}}
\tablehead{
\colhead{Source }    &  \colhead{Number of stars} 
}
\startdata
ISPI detected in \jb-, \hb- and \kb-bands & 4851\\
ISPI detected in \jb- and \hb-bands, not in \kb-band & 92\\
ISPI\  detected in \hb- and \kb-bands, not in \jb-band & 1096\\
ISPI\ detected in \jb- and \kb-bands, not in \hb-band & 85\\
ISPI\ detected only in \jb-band & 175\\
ISPI detected only in \hb-band & 80\\
ISPI detected only in \kb-band & 251\\
ISPI\  diffuse & 933\\
{\hskip 1cm ISPI\ Total} & 7563\\
Extracted from the 2MASS catalog & 174\\
Extracted from the \cite{Mue02} catalog & 22\\
{\hskip 1cm Grand\ Total} & 7759\\
\tableline
ISPI\ detected in 30~s images & 6147\\
 ISPI detected in 3~s images & 1416\\
\tableline
ISPI\  first run detection & 6305\\
ISPI\  hidden companions & 325\\
ISPI\ diffuse & 933\\
\tableline
\enddata
\end{deluxetable}
\clearpage

?

%%%
%TABLE\ 5
%
%\clearpage
\begin{deluxetable}{rrrrrrrrrrrrrrrrrrrrrrrrr}
\tabletypesize{\scriptsize}
\tablecaption{ISPI\ Photometry and Astrometry of the Orion Nebula Cluster\label{tab:samplecat}}
\tablewidth{0pt}
\tablehead{
\colhead{ID} & \colhead{RA} & \colhead{DEC} & \colhead{RAJ} & \colhead{DECJ} &
\colhead{RAH} & \colhead{DECH} & \colhead{RAK} & \colhead{DECK} & \colhead{J} & \colhead{DJ} & \colhead{H} &
\colhead{DH} & \colhead{K} & \colhead{DK} & \colhead{J\_AP} & \colhead{DJ\_AP} & \colhead{H\_AP} &
\colhead{DH\_AP} & \colhead{K\_AP} & \colhead{DK\_AP} & \colhead{IMAGE} & \colhead{STATUS} & \colhead{EXPTIME} & \colhead{CLONE}
}
\startdata
1 & 5.5891618 & -5.2200644 & 0.0000000 & 0.0000000 & 0.0000000 & 0.0000000 & 5.5891618 & -5.2200644 & 0.00000 & 0.00000 & 0.00000 & 0.00000 & 0.00000 & 0.00000 & 0.00000 & 0.00000 & 0.00000 & 0.00000 & 14.7662 & 0.127338 & 1 & 30 & 3 & 0\\
2 & 5.5953268 & -5.2236077 & 5.5953268 & -5.2236077 & 5.5953268 & -5.2236077 & 5.5953268 & -5.2236077 & 0.00000 & 0.00000 & 0.00000 & 0.00000 & 0.00000 & 0.00000 & 18.3755 & 0.165048 & 17.1395 & 0.0658121 & 16.5264 & 0.109725 & 1 & 30 & 3 & 0\\
3 & 5.5892962 & -5.2290140 & 5.5892962 & -5.2290140 & 5.5892962 & -5.2290140 & 5.5892962 & -5.2290140 & 0.00000 & 0.00000 & 0.00000 & 0.00000 & 0.00000 & 0.00000 & 21.6719 & 3.20662 & 18.8478 & 0.312653 & 16.5874 & 0.435113 & 1 & 30 & 3 & 0\\
4 & 5.5957626 & -5.2373337 & 5.5957626 & -5.2373337 & 5.5957626 & -5.2373337 & 5.5957626 & -5.2373337 & 0.00000 & 0.00000 & 0.00000 & 0.00000 & 0.00000 & 0.00000 & 18.2395 & 0.144179 & 16.8748 & 0.0518666 & 16.2750 & 0.0847839 & 1 & 30 & 3 & 0\\
5 & 5.5977591 & -5.2379900 & 5.5977591 & -5.2379900 & 5.5977591 & -5.2379900 & 5.5977591 & -5.2379900 & 0.00000 & 0.00000 & 0.00000 & 0.00000 & 0.00000 & 0.00000 & 14.9843 & 0.0181128 & 14.1678 & 0.0144228 & 13.8639 & 0.0164303 & 1 & 30 & 3 & 0\\
6 & 5.5977757 & -5.2385170 & 5.5977757 & -5.2385170 & 5.5977757 & -5.2385170 & 5.5977757 & -5.2385170 & 0.00000 & 0.00000 & 0.00000 & 0.00000 & 0.00000 & 0.00000 & 16.1198 & 0.0362412 & 15.5781 & 0.0223180 & 15.1644 & 0.0347833 & 1 & 30 & 3 & 0\\
\dots & \dots & \dots & \dots & \dots & \dots & \dots & \dots & \dots & \dots & \dots & \dots & \dots & \dots & \dots & \dots & \dots & \dots & \dots & \dots & \dots & \dots & \dots & \dots & \dots \\
50 & 5.5892075 & -5.3059220 & 5.5892064 & -5.3059297 & 5.5892100 & -5.3059316 & 5.5892075 & -5.3059220 & 13.0184 & 0.0607901 & 11.9386 & 0.0319145 & 11.3727 & 0.0440246 & 12.8548 & 0.0307046 & 11.8450 & 0.0206606 & 11.3464 & 0.0269704 & 1 & 3 & 1 & 50\\
51 & 5.5884842 & -5.3057194 & 5.5884816 & -5.3056698 & 5.5884806 & -5.3056545 & 5.5884842 & -5.3057194 & 12.7213 & 0.0437424 & 11.9917 & 0.0212242 & 11.6676 & 0.0364081 & 12.6560 & 0.0289495 & 11.9254 & 0.0186968 & 11.5759 & 0.0257633 & 1 & 3 & 1 & 51\\
52 & 5.5919210 & -5.3049874 & 5.5919174 & -5.3049865 & 5.5919189 & -5.3049850 & 5.5919210 & -5.3049874 & 11.4727 & 0.0458089 & 10.7665 & 0.0297185 & 10.5796 & 0.0417144 & 11.4616 & 0.0245957 & 10.7047 & 0.0164845 & 10.5815 & 0.0225129 & 1 & 3 & 1 & 52\\
53 & 5.5919566 & -5.3021426 & 5.5919551 & -5.3021808 & 5.5919540 & -5.3021369 & 5.5919566 & -5.3021426 & 17.4306 & 0.218837 & 16.6371 & 0.107383 & 16.3148 & 0.208574 & 17.3613 & 0.256974 & 16.5880 & 0.123214 & 16.2403 & 0.230356 & 1 & 3 & 1 & 53\\
\dots & \dots & \dots & \dots & \dots & \dots & \dots & \dots & \dots & \dots & \dots & \dots & \dots & \dots & \dots & \dots & \dots & \dots & \dots & \dots & \dots & \dots & \dots & \dots & \dots \\
3597 & 5.5816976 & -5.4131470 & 5.5817022 & -5.4131479 & 5.5816945 & -5.4131670 & 5.5816976 & -5.4131470 & 16.7388 & 0.112858 & 16.0862 & 0.0527115 & 15.5653 & 0.0936435 & 0.00000 & 0.00000 & 0.00000 & 0.00000 & 0.00000 & 0.00000 & 5 & 30 & 2 & 3597\\
\dots & \dots & \dots & \dots & \dots & \dots & \dots & \dots & \dots & \dots & \dots & \dots & \dots & \dots & \dots & \dots & \dots & \dots & \dots & \dots & \dots & \dots & \dots & \dots & \dots \\
3678 & 5.5823130 & -5.4152808 & 0.0000000 & 0.0000000 & 5.5823110 & -5.4152832 & 5.5823130 & -5.4152808 & 0.00000 & 0.00000 & 18.0264 & 0.141975 & 17.3747 & 0.147156 & 0.00000 & 0.00000 & 0.00000 & 0.00000 & 0.00000 & 0.00000 & 5 & 30 & 2 & 3678\\
\dots & \dots & \dots & \dots & \dots & \dots & \dots & \dots & \dots & \dots & \dots & \dots & \dots & \dots & \dots & \dots & \dots & \dots & \dots & \dots & \dots & \dots & \dots & \dots & \dots \\
4109 & 5.5791845 & -5.3399510 & 5.5791916 & -5.3398418 & 5.5791794 & -5.3398919 & 5.5791845 & -5.3399510 & 17.3715 & 0.134945 & 16.3929 & 0.0630108 & 15.7189 & 0.104973 & 0.00000 & 0.00000 & 0.00000 & 0.00000 & 0.00000 & 0.00000 & 5 & 30 & 2 & 4109\\
\dots & \dots & \dots & \dots & \dots & \dots & \dots & \dots & \dots & \dots & \dots & \dots & \dots & \dots & \dots & \dots & \dots & \dots & \dots & \dots & \dots & \dots & \dots & \dots & \dots \\
82996 & 5.5871119 & -5.5159988 & 0.0000000 & 0.0000000 & 0.0000000 & 0.0000000 & 0.0000000 & 0.0000000 & 10.1820 & 0.0230000 & 9.26200 & 0.0300000 & 8.62400 & 0.0190000 & 0.00000 & 0.00000 & 0.00000 & 0.00000 & 0.00000 & 0.00000 & 8 & 0 & 0 & 82996\\
\dots & \dots & \dots & \dots & \dots & \dots & \dots & \dots & \dots & \dots & \dots & \dots & \dots & \dots & \dots & \dots & \dots & \dots & \dots & \dots & \dots & \dots & \dots & \dots & \dots \\
82620 & 5.5860662 & -5.4647789 & 0.0000000 & 0.0000000 & 0.0000000 & 0.0000000 & 0.0000000 & 0.0000000 & 7.74300 & 0.0240000 & 7.63600 & 0.0470000 & 7.47200 & 0.0200000 & 0.00000 & 0.00000 & 0.00000 & 0.00000 & 0.00000 & 0.00000 & 8 & 0 & 0 & -52620\\
\dots & \dots & \dots & \dots & \dots & \dots & \dots & \dots & \dots & \dots & \dots & \dots & \dots & \dots & \dots & \dots & \dots & \dots & \dots & \dots & \dots & \dots & \dots & \dots & \dots \\
81916 & 5.5811772 & -5.5523648 & 0.0000000 & 0.0000000 & 0.0000000 & 0.0000000 & 0.0000000 & 0.0000000 & 8.45700 & 0.0320000 & 7.95600 & 0.0510000 & 7.85200 & 0.0240000 & 0.00000 & 0.00000 & 0.00000 & 0.00000 & 0.00000 & 0.00000 & 8 & 0 & 0 & 81916\\
\dots & \dots & \dots & \dots & \dots & \dots & \dots & \dots & \dots & \dots & \dots & \dots & \dots & \dots & \dots & \dots & \dots & \dots & \dots & \dots & \dots & \dots & \dots & \dots & \dots \\
81767 & 5.5797176 & -5.5707202 & 0.0000000 & 0.0000000 & 0.0000000 & 0.0000000 & 0.0000000 & 0.0000000 & 7.22100 & 0.0190000 & 6.96400 & 0.0340000 & 6.64100 & 0.0180000 & 0.00000 & 0.00000 & 0.00000 & 0.00000 & 0.00000 & 0.00000 & 8 & 0 & 0 & 81767\\
\dots & \dots & \dots & \dots & \dots & \dots & \dots & \dots & \dots & \dots & \dots & \dots & \dots & \dots & \dots & \dots & \dots & \dots & \dots & \dots & \dots & \dots & \dots & \dots & \dots \\
\enddata
%% Text for table notes should follow after the \enddata but before
%% the \end{deluxetable}. Make sure there is at least one \tablenotemark
%% in the table for each \tablenotetext.
\tablecomments{Table \ref{tab:samplecat} is published in its entirety in the electronic edition. A portion is shown here for guidance regarding its form and content. See Appendix A for a detailed description of the fields.}
%\tablenotetext{a}{Sample footnote for table~\ref{tbl-1} that was generated with the deluxetable environment}
%\tablenotetext{b}{Another sample footnote for table~\ref{tbl-1}}
\end{deluxetable}
\clearpage


\begin{thebibliography}{}
          \bibitem[Baraffe et al.(2002)]{2002A&A...382..563B} Baraffe, I., Chabrier, G., Allard, F., \& Hauschildt, P.~H.\ 2002, \aap, 382, 563
\bibitem[Baraffe et al.(1998)]{BCAH98} Baraffe, I., Chabrier, G., Allard, F., \& Hauschildt, P.~H.\ 1998, \aap, 337, 403 %GAETANO SCANDARIATO
\bibitem[Bertin et al.(2002)]{2002ASPC..281..228B} Bertin, E., Mellier, Y., Radovich, M., Missonnier, G., Didelon, P., \& Morin, B.\ 2002, Astronomical Data Analysis Software and Systems XI, 281, 228
\bibitem[Cardelli et al.(1989)]{Cardelli89} Cardelli, J.~A., Clayton, G.~C., \& Mathis, J.~S.\ 1989, \apj, 345, 245
\bibitem[Carpenter(2001)]{Carp01} Carpenter, J.~M.\ 2001, \aj, 121, 2851
\bibitem[Chabrier et al.(2000)]{CBAH00} Chabrier, G., Baraffe, I., Allard, F., \& Hauschildt, P.\ 2000, \apj, 542, 464 %GAETANO SCANDARIATO
\bibitem[Da Rio et al.(2009)]{DaRio+09}Da Rio, N., Robberto, M., 
Soderblom, D. R., Panagia,  N., Hillenbrand, L. A., Palla, F., and Stassun, K. 2009, \apjs, in press
\bibitem[de Grijs et al.(2002a)]{Gra} de Grijs, R., Gilmore, G.~F., Mackey, A.~D., Wilkinson, M.~I., Beaulieu, S.~F., Johnson, R.~A., \& Santiago, B.~X.\ 2002, \mnras, 337, 597  %GAETANO SCANDARIATO
\bibitem[de Grijs et al.(2002b)]{Grb} de Grijs, R., Gilmore, G.~F., Johnson, R.~A., \& Mackey, A.~D.\ 2002, \mnras, 331, 245 %GAETANO SCANDARIATO
\bibitem[de Grijs et al.(2002c)]{Grc} de Grijs, R., Johnson, R.~A., Gilmore, G.~F., \& Frayn, C.~M.\ 2002, \mnras, 331, 228 %GAETANO SCANDARIATO
\bibitem[Goldsmith et al.(1997)]{Goldsmith+97} Goldsmith, P.~F., Bergin, E.~A., \& Lis, D.~C.\ 1997, \apj, 491, 615 
\bibitem[Hillenbrand(1997)]{LAH97} Hillenbrand, L.~A.\ 1997, \aj, 113, 1733 %GAETANO SCANDARIATO
\bibitem[Hillenbrand \& Carpenter(2000)]{HC00} Hillenbrand, L.~A., \& Carpenter, J.~M.\ 2000, \apj, 540, 236 
\bibitem[Hillenbrand \& Hartmann(1998)]{HH98} Hillenbrand, L.~A., \& Hartmann, L.W. 1998, \apj, 492, 540
\bibitem[Lada et al.(2008)]{Lada08} Lada, C.~J., Muench, A.~A., Rathborne, J., Alves, J.~F., \& Lombardi, M.\ 2008, \apj, 672, 410 
\bibitem[Landolt (1992)]{landolt92} Landolt, A.\ U.\ 1992, \aj, 104, 340 %BILL SHERRY
\bibitem[Lee et al.(2005)]{Lee05} Lee, H.-T., Chen, W.~P., Zhang, Z.-W., \& Hu, J.-Y.\ 2005, \apj, 624, 808 
\bibitem[Lucas \& Roche(2000)]{LR00} Lucas, P.~W., \&\ Roche, P.~F.,  2000, \mnras, 205, 361
\bibitem[Lucas et al.(2005)]{LR05} Lucas, P.~W., Roche, P.~F., \& Tamura, M.\ 2005, \mnras, 361, 211 %GAETANO SCANDARIATO
\bibitem[Menten et al.(2007)]{Menten07} Menten, K.~M., Reid, M.~J., Forbrich, J., \& Brunthaler, A.\ 2007, \aap, 474, 515 
\bibitem[Meyer et al.(1997)]{Meyer97} Meyer, M.~R., Calvet, N., \& Hillenbrand, L.~A.\ 1997, \aj, 114, 288 
\bibitem[Mink (1997)]{mink97} Mink, D.\ J.\, 1997, in {\em Astronomical Data Analysis Software and Systems VI}, A.\ S.\ P.\ Conference Series, Vol.\ 125, G.\ Hunt and H.\ E.\ Payne, eds.\, p.\ 249%BILL SHERRY
\bibitem[Muench et al.(2002)]{Mue02} Muench, A.~A., Lada, E.~A., Lada, C.~J., \& Alves, J.\ 2002, \apj, 573, 366 %GAETANO SCANDARIATO
\bibitem[Muench et al.(2008)]{Muench+08} Muench, A., Getman, K., Hillenbrand, L., Preibisch, T. 2008, in ``Handbook of Star Forming Regions, Volume I: The Northern Sky'' ASP Monograph Publications, Vol. 4. (Editor B. Reipurth), 483
%\bibitem[Persson et al.(1998)]{Persson+2003} Persson S. E. et al., 2003, \aj, 116, 2475

\bibitem[O'Dell (2008)]{ODell+08} O'Dell C. R., Muench, A., Smith, N., Zapata, L. 2008, in ``Handbook of Star Forming Regions, Volume I: The Northern Sky ASP Monograph Publications'', Vol. 4.  (Editor B. Reipurth), 544
\bibitem[O'Dell et al.(2009)]{ODell+09}O'Dell, c. R., Henney, W. J., Abel, N. P., Ferland, G. J., and Arthur, S. J. 2009, \aj 137, 367
\bibitem[Persson et al.(1998)]{Persson+1998} Persson, S.~E., Murphy, D.~C., Krzeminski, W., Roth, M., \& Rieke, M.~J.\ 1998, \aj, 116, 2475 
\bibitem[\protect\citeauthoryear{Press et al.}{1997}]{PressC} Press W.~H., Teukolsky S.~A., Vetterling W.~T., Flannery B.~P. 1997, Cambridge University Press %GAETANO SCANDARIATO
\bibitem[Rieke \& Lebofsky(1985)]{Rie85} Rieke, G.~H., \& Lebofsky, M.~J.\ 1985, \apj, 288, 618
\bibitem[Robberto, Beckwith \&\ Panagia(2002)]{RBP02}Robberto, M., Beckwith, S.~V.~W., \& Panagia, N.\ 2002, \apj, 578, 897
\bibitem[Scandariato (2008)]{Sca08} Scandariato, G. 2008, Master Thesis, University of Catania%GAETANO SCANDARIATO
\bibitem[Sherry (2003)]{thesis} Sherry, W.\ H.\ 2003, {\em The Young Low-Mass Population of Orion's Belt}, Ph.\ D.\ Thesis, State University of New York, Stony Brook%BILL SHERRY
\bibitem[Siess et al.(2000)]{Sie00} Siess, L., Dufour, E., \& Forestini, M.\ 2000, \aap, 358, 593
\bibitem[Stetson(1987)]{stetson87} Stetson, P.~B.\ 1987, \pasp, 99, 191
\bibitem[Xin-yue et al.(2009)]{Xin-yue+09}
Xin-yue, E, Zhi-bo, J., Yan-ning, F. , 2009, Chinese Astronomy and Astrophysics, 33, 139
\end{thebibliography}
\end{document}